\documentclass{vgtc}                          %

\graphicspath{{figures/}{pictures/}{images/}{./}} %

\usepackage{times}                     %

\usepackage{tabu}                      %
\usepackage{booktabs}                  %
\usepackage{lipsum}                    %
\usepackage{mwe}                       %
\usepackage{textcomp}
\usepackage[T1]{fontenc}
\usepackage{mathptmx}                  %

\usepackage[per-mode=symbol]{siunitx}
\onlineid{2391}

\vgtccategory{Methodological papers}

\vgtcinsertpkg

\title{Comparing Vibrotactile and Skin-Stretch Haptic Feedback for Conveying Spatial Information of Virtual Objects to Blind VR Users}

\author{
Jiasheng Li\thanks{e-mail: jsli@umd.edu}\\ %
    \scriptsize University of Maryland, College Park
\and
Zining Zhang\thanks{e-mail: znzhang@umd.edu}\\ %
    \scriptsize University of Maryland, College Park
\and
Zeyu Yan\thanks{e-mail: zeyuy@umd.edu}\\ %
    \scriptsize University of Maryland, College Park
\and    
Yuhang Zhao\thanks{e-mail: yuhang.zhao@cs.wisc.edu}\\ %
    \scriptsize University of Wisconsin-Madison
\and    
Huaishu Peng\thanks{e-mail: huaishu@umd.edu}\\ %
    \scriptsize University of Maryland, College Park
}

\abstract{
Perceiving spatial information of a virtual object (e.g., direction, distance) is critical yet challenging for blind users seeking an immersive virtual reality (VR) experience.
To facilitate VR accessibility for blind users, in this paper, we investigate the effectiveness of two types of haptic cues---vibrotactile and skin-stretch cues---in conveying the spatial information of a virtual object when applied to the dorsal side of a blind user's hand. 
We conducted a user study with 10 blind users to investigate how they perceive static and moving objects in VR with a custom-made haptic apparatus.
Our results reveal that blind users can more accurately understand an object's location and movement when receiving skin-stretch cues, as opposed to vibrotactile cues.
We discuss the pros and cons of both types of haptic cues and conclude with design recommendations for future haptic solutions for VR accessibility.
} %


\begin{document}
\maketitle

\section{Introduction}
Spatial information is essential for users to perceive and navigate a 3D virtual reality (VR) environment. 
For sighted users, most spatial cues---such as the relative positions and directions of static objects, or the motion of moving objects---are delivered primarily through visual simulations on head-mounted displays (HMDs).
Unfortunately, this vision-centric approach poses significant barriers to blind users, preventing them from perceiving and understanding spatial information in a VR space. 
As a result, emerging VR user interfaces risk excluding the 1.3 billion blind and low-vision individuals worldwide \cite{accessible_design}, as highlighted by Dr. Azenkot in her keynote talk at IEEE VR 2024~\cite{azenkot2024xr}.

One area of focus for making VR accessible has been assistive auditory systems that provide blind users with spatial information (e.g., ~\cite{blauert1997spatial,collignon2009cross,gougoux2005functional,collignon2011functional,renier2010preserved}).
However, while effective, audio-only feedback often falls prey to interference from various sound sources, 
such as ambient or environmental noises, background music, or speech from other avatars sharing the same virtual space~\cite{play_blind, zelda}.
As a result, relying exclusively on auditory feedback is frequently inadequate for delivering clear and reliable spatial information.

Haptics, as an independent sensory channel, offers the potential to convey additional spatial information without disrupting the auditory cues in a virtual environment. While various haptic mechanisms have been extensively explored for VR applications, most have not been designed to assist blind users in understanding spatial information within VR. For example, one-dimensional vibration is the de facto standard in commercial VR controllers. Beyond this, haptic gloves with multiple vibrotactile motors (e.g., \cite{TactileGlove, haptiX}) and phantom-sensation-based wearables (e.g., \cite{phantom_senstion_wearables}) have also been explored to provide haptic feedback in VR environments. More recently, skin-stretch mechanisms (e.g.,~\cite{skin_stretch_on_face, drag_tap_vib, skin_stretch_on_leg}) have been incorporated into VR headset hardware~\cite{skin_stretch_on_face}, enabling the rendering of two-dimensional sensations. However, these devices primarily aim to enhance VR immersiveness rather than provide spatial information, and they have not been evaluated with the blind and low-vision community. As a result, the use of haptics to support the comprehension of spatial information in VR remains underexplored.

In this paper, we present an empirical study investigating how haptic mechanisms can assist blind users in understanding spatial information rendered in VR. Considering the various haptic mechanisms (e.g., \cite{3dSystemsPhantomPremium, TactoRing, lips_non-contact_tactile}) and body locations (e.g., the forehead, torso) where haptic stimuli can be applied, the design space is vast. However, in this paper, we are particularly interested in two haptic mechanisms---on-skin vibration and skin-stretch---and their effectiveness in enhancing blind users' comprehension of \textit{static information} (e.g., the location and direction of a virtual table) and \textit{dynamic information} (e.g., the trajectory of an approaching avatar or a flying object in a VR game) when applied to \textit{the dorsal side of the hand}.
We chose this combination of haptic feedback mechanisms and body location for three primary design considerations. First, previous research has showed that the skin on the hand has relatively high sensitivity compared to many other body locations~\cite{gallace2009cognitive}, making it potentially effective for rendering spatial information. Second, compared to other parts of the hand, the dorsal side offers a sufficiently flat area that is not involved in VR operations (unlike the palm and fingers, which are essential for holding and operating VR controllers), making it a practical location for haptics stimuli. Third, both vibration and skin-stretch mechanisms can be engineered with small form factors and low cost, critical consideration when designing assistive technology~\cite{low-cost_disability_device}.

Our user study involved 10 blind participants in a within-subject experimental design. We used a custom apparatus---a three-gantry, desktop-grounded experimental device---to render both haptic cues on the dorsal side of the hand (Figure~\ref{fig:haptic_device}). Unlike a wearable setup (e.g., a glove), our apparatus adapts to users with different hand sizes and removes the need for repeatedly donning and doffing devices when transitioning between different haptic mechanisms. During the study, different locations and movement patterns were applied to the dorsal side of each participant's hand, and participants were asked to describe the direction or the shape they perceived. 

The results indicated that while blind users were capable of discerning an object's location and movement patterns through both types of haptic feedback, the skin-stretch mechanism enabled more accurate perception than vibrotactile feedback. 
On the other hand, vibrotactile cues---being the most frequently encountered  haptic mechanism among blind users---provided clear sensations and were preferred by several participants. 
We discuss the advantages and disadvantages of these haptic mechanisms and offer design recommendations for future VR systems to better accommodate blind users.

\section{Related work}

\subsection{Spatial Information Through Vibration}\label{haptics}

While one-dimensional vibration is a common method for delivering haptic cues in modern VR devices, it is primarily used to enhance immersion rather than to convey spatial information. To render spatial information, researchers have investigated the use of multiple vibrotactile motors across various body parts, such as forearms~\cite{weber2011evaluation}, head~\cite{on_head_vib_display}, waist~\cite{tsukada2004activebelt}, hands~\cite{haptiX,dynamic_tactile, TactileGlove}, and back~\cite{tan2003haptic}, among others (see \cite{vibrotactile_survey} for a comprehensive survey). Recently, work such as HaptiX~\cite{haptiX} demonstrates that the dorsal side of the hand can serve as a viable area for receiving directional cues via a multi-vibration haptic glove. However, these studies exclusively focused on sighted users and their perceptions.

A few notable works have explored the use of multiple vibration motors to provide spatial information to blind users. SpaceSense~\cite{spaceSense}, for example, created a three-by-three grid of vibrotactile motors on the backside of a smartphone. When blind users hold the phone, vibrations from the various motors provide directional guidance for wayfinding. Hong et al.~\cite{hong2017evaluating} investigated a similar wayfinding task using a wristband with four to eight motors, while Zhao et al.~\cite{vibhand} placed four vibration motors on the back of the hand to guide blind users in exploring visualizations printed on paper.

In our work, we also focus on haptic rendering specifically for blind users. Unlike previous work, we aim to provide a more systematic understanding of how blind users perceive the dorsal side of the hand as a potential location for receiving up to eight directional cues. We also explored two different types of information: static information as directional cues and dynamic information in the form of trajectory shapes.

\subsection{Skin-Stretch Mechanism and Moving Stimuli}

Besides vibration, the skin-stretch mechanism is another promising haptic rendering method considered in this work, given its two-dimensional haptic rendering capability and its potential to be engineered into small form factors (e.g., ~\cite{ion}).
The underlining principle of skin-stretch devices is to deliver haptic feedback by directly pressing and stretching the user’s skin~\cite{hayward2000tactile}. 
As such mechanisms usually require less physical movement, they have recently been integrated into various wearable devices, applied to the legs~\cite{skin_stretch_on_leg, kent2023biomechanically}, forearms~\cite{shim, bardot, bark2010rotational}, wrists~\cite{ion}, finger joints~\cite{je_design, TactoRing, SpinOcchio}, tips~\cite{schorr2013sensory}, and even the face~\cite{skin_stretch_on_face, miyakami2022head}.

While existing research has underscored the potential of skin-stretch mechanisms in providing rich haptic information across various applications, similar to research in multi-array vibrations, the bulk of these studies has primarily focused on sighted users.
The perception of skin-stretch by blind individuals, and its efficacy in conveying spatial information within a VR setup, remains unexplored. 

Thus, the primary goal of our work is to gain an in-depth understanding of vibration and skin-stretch mechanisms for conveying spatial information to blind users, with the intention of integrating them into future accessible VR devices. We believe this understanding is crucial before undertaking meaningful system engineering research. As technical researchers, we strive to avoid creating ``disability dongles'' due to a lack of empirical insight, as strongly emphasized by accessibility researchers and advocates~\cite{leslie_2022_disabilitydongle}.

\section{Method}

\begin{table*}[]
\centering
\caption{Demographic information of the 10 blind participants.}
\label{tab:participants}
\begin{tabular}{p{35pt}p{50pt}p{70pt}p{70pt}p{70pt}p{75pt}}
\toprule
\textbf{ID} &
\textbf{Gender} &
\multicolumn{1}{l}{\textbf{Vision Level}} &
\multicolumn{1}{l}{\textbf{Age Range}} &
\multicolumn{1}{l}{\textbf{Dominant Hand}} &
\multicolumn{1}{l}{\textbf{\begin{tabular}[c]{@{}c@{}}Self Rated \\ Experience in VR\end{tabular}}}\\ 
\midrule

P1 &  Female &  Blind &  30-40 &  Right &  Minimal experience \\
P2 &  Female &  Blind &  over 60 &  Right &  No experience  \\
P3 &  Female &  Blind &  40-50 &  Right &  No experience \\
P4 &  Female &  Blind &  30-40 &  Right &  No experience \\
P5 &  Female &  Blind &  50-60 &  Right &  Minimal experience \\
P6 &  Female &  Blind &  40-50 &  Right &  No experience \\
P7 &  Female &  Blind &  50-60 &  Right &  No experience \\
P8 &  Female &  Blind &  50-60 &  Right &  No experience \\
P9 &  Female &  Blind &  18-20 &  Right &  Minimal experience \\
P10 &  Male &  Blind &  30-40 &  Right &  No experience \\ \bottomrule
\end{tabular}%
\end{table*}

We aim to investigate how vibrations and skin-stretch mechanisms can assist blind users in perceiving spatial information in VR. 
Specifically, our focus is on two common types of spatial information: (1) the distance and directional information of a static virtual object, for example, a virtual desk located to the north or south relative to the user's current standing position; and (2) the trajectory of a moving virtual object of interest, such as an avatar approaching from behind or the flying path of a virtual frisbee in a VR game.

We conducted a within-subject, single-session study comprising two tasks, lasting approximately \SIrange{120}{150}{\minute}. 
Ten blind participants were recruited for the study. 
They were compensated at a rate of \SI{30}{USD} per hour in cash. 
The study was recorded in video and audio for data analysis purposes and was reviewed and approved by the University Institutional Review Board (IRB).

\subsection{Participants}\label{participants}
We recruited ten participants (Table \ref{tab:participants}) through the mailing list of the National Federation of the Blind (NFB) organization (9 females, 1 male). 
Due to the use of convenience sampling, we were unable to control for age and sex.
All participants self-reported as blind and indicated that their dominant hand was the right hand. 
We collected basic background information on participants' experience with VR devices and their familiarity with haptic feedback. 
Additionally, we asked participants how they typically describe directional information or object's location in their daily lives.

All participants had little to no experience with VR. 
Only three participants mentioned having minimal VR experience, which was limited to wearing a head-mounted display or hearing audio from the menu screen without fully understanding the virtual environment. All participants were familiar with vibrotactile feedback, commonly used in devices such as smartphones and smartwatches for notifications. 
None had prior experience with skin-stretch devices. 
Participants reported a preference for using relative directions (e.g., left, right, forward, backward) over cardinal directions (e.g., north, south) to describe their spatial relationship with surrounding objects.

\subsection{Apparatus}\label{apparatus}
\begin{figure}[h]
    \includegraphics[width=\columnwidth]{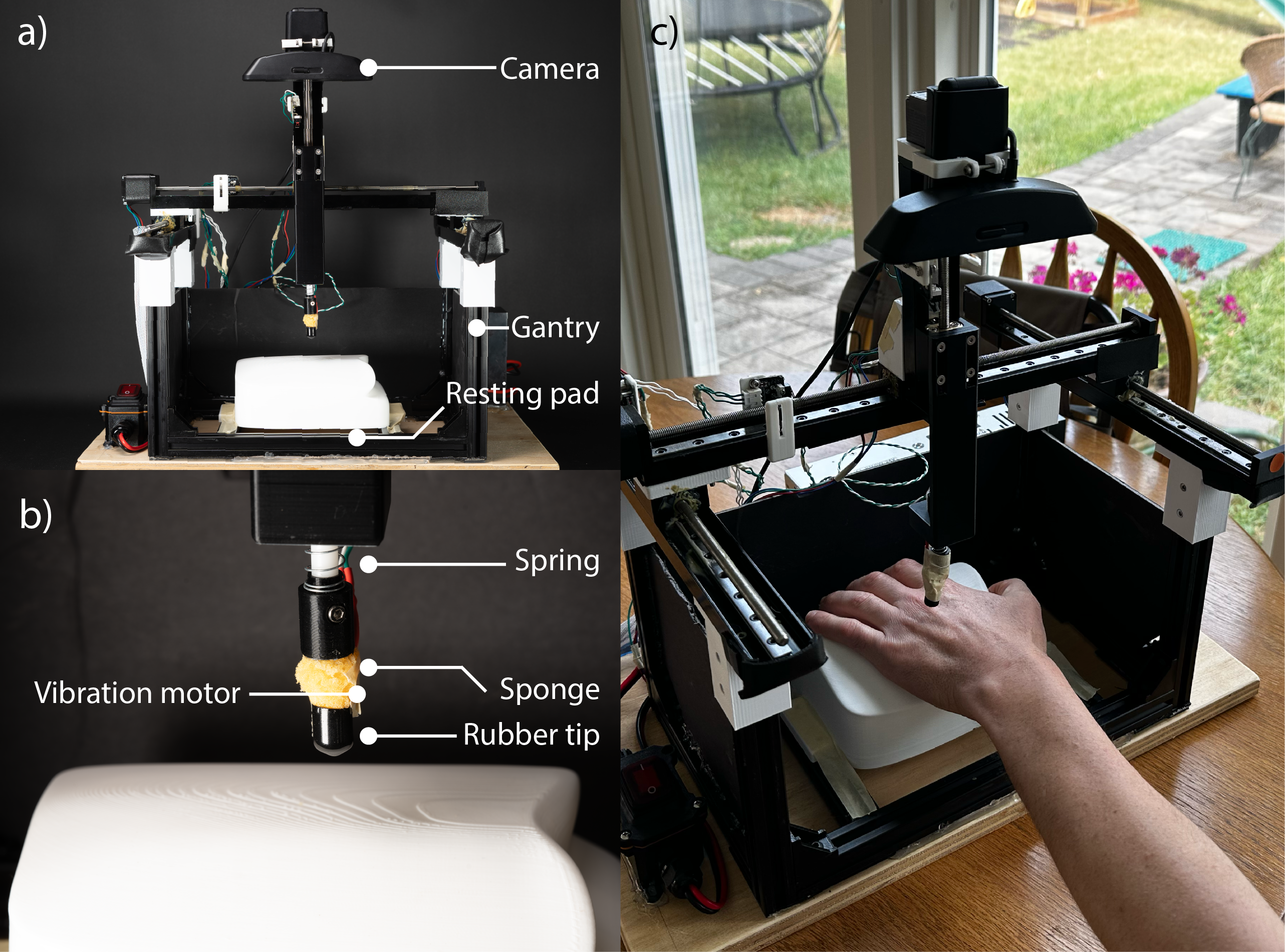}
    \caption{Haptic device setup: a) Front view of the three-axis gantry system haptic device, b) the touch probe, and c) study conducted at a participant’s home.}
    \label{fig:haptic_device}
\end{figure}

We designed a three-gantry experimental device \footnote{The experimental device is open-sourced and available at \url{https://github.com/jsli96/handHapticforBlind}} for our study, as illustrated in Figure~\ref{fig:haptic_device}.
Since the main focus of our research was to advance the understanding of how different haptic mechanisms provide spatial information for blind users, rather than specific engineering efforts in wearable integration, we did not aim for a minimized form factor. 
Instead, our setup prioritizes accommodating various hand sizes, enabling both multi-location vibration and skin-stretch within the same setup to ensure compatibility and reduce the effort required to switch devices, while also ensuring participant safety.

As shown in Figure~\ref{fig:haptic_device}, the three-axis gantry system is equipped with a multifunctional touch probe capable of rendering both skin-stretch and vibrotactile feedback. The touch probe can move in the X, Y, and Z directions within the device's frame, with motion ranges of \SI{300}{\milli\meter}, \SI{250}{\milli\meter}, and \SI{100}{\milli\meter}, respectively. These ranges are sufficient to cover the size of an adult's hand for haptic rendering.
A rubber touch probe is attached to the gantry system via a spring and an additional sponge. Both the spring and the sponge ensure that the experimental device applies moderate force (less than \SI{2}{\newton}) to the user's hand without causing discomfort. A vibration motor is attached to the touch probe to provide vibration feedback.
To ensure consistent hand positioning and comfort throughout the testing procedure, a 3D-printed resting pad is located at the bottom of the device, where participants rest their non-dominant hand. Lastly, a safety switch is installed next to the experimental device, allowing for an immediate power cutoff if necessary.

\subsubsection{Calibration}
To account for variations in participants' hand sizes and the slight curvature of the dorsal side of the hand, a camera was mounted on top of the gantry frame, facing downward, and a calibration process was conducted for each participant prior to the study.
The process ensured consistent touch rendering at the same position across different user's hands and guaranteed that the touch probe was always in good contact with the skin for each trial.

The calibration process is as follows:
Once a participant places their hand on the pad, the device's camera captures a photo to calculate the center point of the dorsum of the hand using MediaPipe~\cite{mediapipe}. 
Next, the height differences are measured at four points along the edges of the user's hand. 
For each point, the device's touch probe is gradually lowered until it makes comfortable yet clear contact with the user's skin. 
The heights of these four points, along with that of the center point, are recorded. 
This height data is then used to render haptic feedback uniformly across the skin by interpolating the Z-axis values from these five points.

\subsubsection{Safety measures}\label{safety}
In addition to the sponge- and spring-equipped touch probe, we implemented three additional measures to ensure the safety of all blind participants during the study. 
First, we limited the speed of the touch probe motion to \SI{8}{\milli\meter\per\second} to provide participants with sufficient reaction time in case they experience any discomfort during the study (which never occurred). 
Additionally, we manually set limits on the maximum displacement of the Z-axis extension arm, to ensure that users can safely withdraw their hand at any time.
Lastly, a safety switch was installed next to the experimental device, allowing for immediate power cutoff when necessary.

It is worth noting that the speed of the touch probe motion can influence participants' spatial perception. Our choice of \SI{8}{\milli\meter\per\second} serves as a safety measure for participants but is also grounded in prior haptic studies. For example, similar speeds have been employed in skin-stretch mechanisms in previous work (e.g.,\cite{doherty2024hapt}). For the vibration condition, while a speed of \SI{8}{\milli\meter\per\second} may result in a \SIrange{2}{3}{\second} interval between vibration locations, previous research has shown that individuals can recall the location of a tactile stimulus for up to \SI{10}{\second}~\cite{lederman2011tactile}, making a \SIrange{2}{3}{\second} interval well within this range.

\subsection{Coordinates on the Dorsum of the Hand}
\begin{figure}[h]
    \includegraphics[width=\columnwidth]{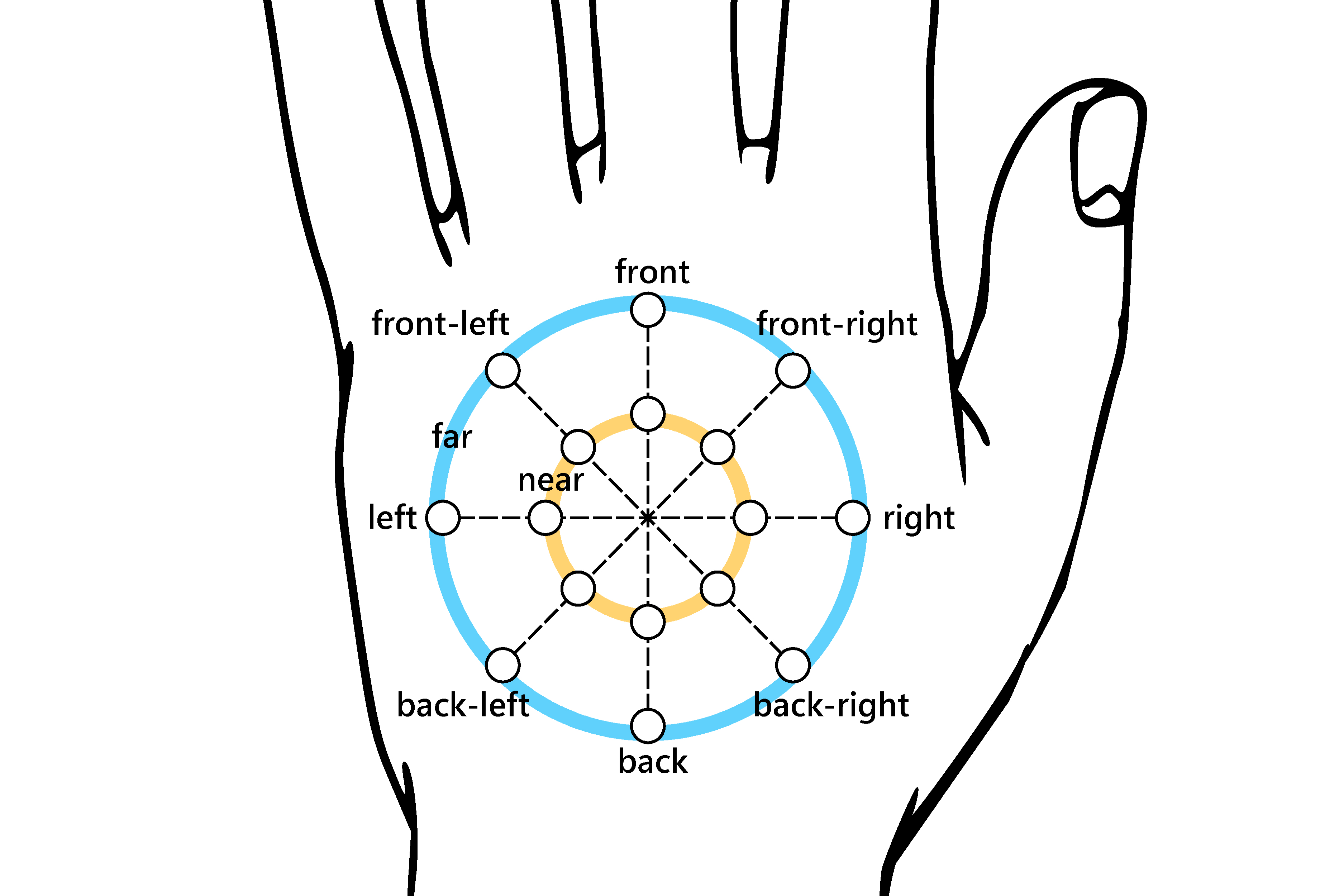}
    \caption{Coordinates and resolution on the dorsum of the hand. The coordinates comprise eight directions and two distances.}
    \label{fig:points_design}
\end{figure}

To map the location of a virtual object relative to the user's position, we designed a coordinate system on the dorsal side of the hand.
Drawing inspiration from previous literature, specifically TactileGlove~\cite{TactileGlove}, we partitioned the surface of the dorsal hand into eight directions, each covering an  angle of \SI{45}{\degree}. 
Since blind individuals often rely on self-referenced spatial coordinate~\cite{blalock2010encoding, SCHMIDT201343, iachini2014does} and are accustomed to using terms such as front, back, left, and right to describe directions (See section~\ref{participants}), we designated the eight directions as front, back, left, right, front-left, front-right, back-left, and back-right (Figure \ref{fig:points_design}).

To represent varying distances, we added two haptic points along each direction, corresponding to near and far objects.
Specifically, points located near the edge of the hand indicate far distance, suggesting that the object is beyond a reachable distance in the VR space, while points closer to the center of the hand represent near distances, indicating that the object is within a reachable distance in VR. The distance between the far and near haptic points is averaged at \SI{10}{\milli\meter} from center to center, with slight variations based on the participant's hand size. This distance-by-area design is inspired by recent VR accessibility research, such as VRBubble~\cite{vrbubble}. 
In total, our proposed coordinate system comprises 16 haptic points and maximizes the use of the dorsal hand area.

\subsection{Rendering Spatial Information}\label{haptic pattern design}
To convey spatial information of a virtual object, we mapped its direction and distance to the aforementioned coordinate system. 
Specifically, the center area of the coordinate system represents the user's current position in the virtual environment. 
The 16 points correspond to the relative direction and distance from the user. 
For example, if a virtual object is located in front of the user and within a reachable distance, this object's location will be mapped onto the hand coordinate in the front direction at the near distance point.
We next detail how we use skin-stretch and vibrotactile cues to render such information.

\begin{figure}[h]
    \includegraphics[width=\columnwidth]{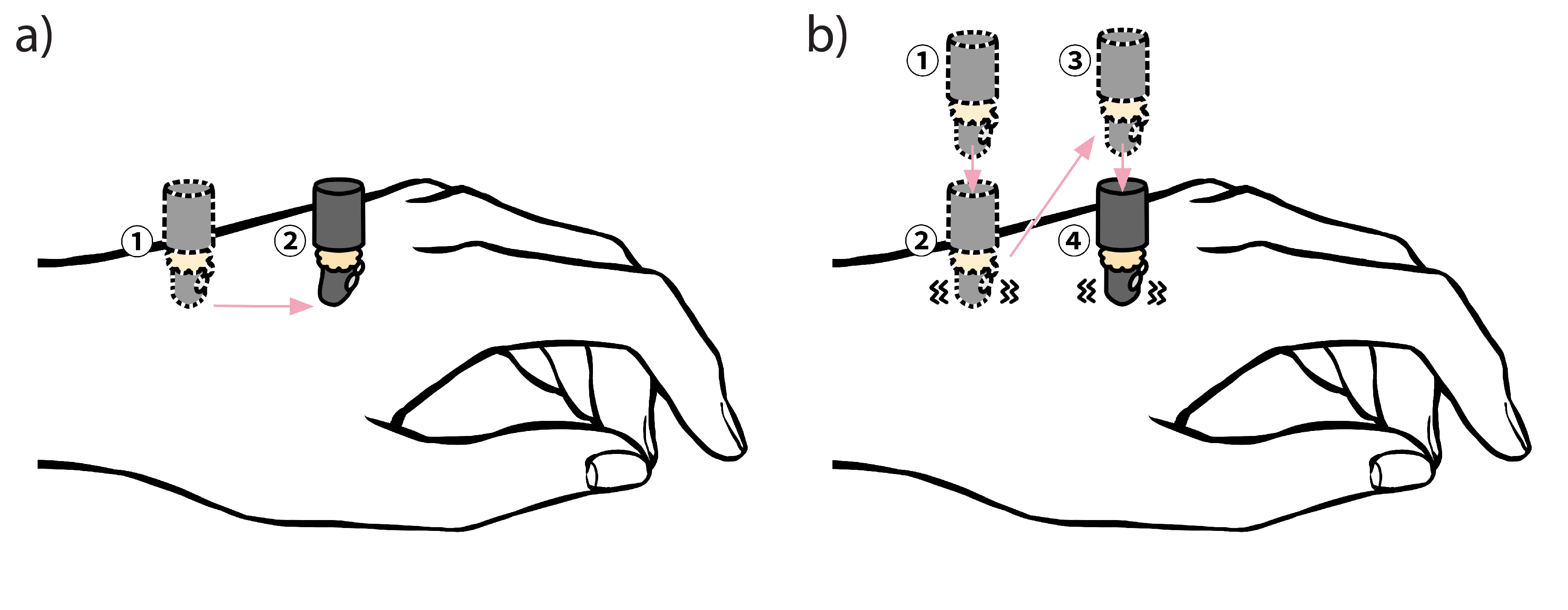}
    \caption{a) The movement of the touch probe in skin-stretch mode. b) The movement of the touch probe in vibration mode.}
    \label{fig:vib_skin_move}
\end{figure}

\subsubsection{Render spatial information through skin-stretch}

To render the location of a static object with skin-stretch, our device draws a straight line on the dorsal side of the hand from its center to the corresponding point in the coordinate system. Specifically, the touch probe first lowers to make contact with the center of the user's hand, then moves across the skin, continuously pressing down to create skin-stretch haptics. Once it reaches the target point, the touch probe lifts up (Figure \ref{fig:vib_skin_move}a). The speed of the motion is \SI{8}{\milli\meter\per\second}, as outlined in Section \ref{safety}.

To render the moving path of a dynamic object on the back of the hand, we map its starting and ending points to the 16-point coordinate system, along with the trajectories of any midpoint lying along the moving path. The touch probe then follows this multi-point path across the skin to represent the movement of the virtual object (Figure \ref{fig:dynamic_patterns}).

\subsubsection{Render spatial information with vibration}
Unlike the skin-stretch mechanism, which drags the skin surface, the touch probe generates vibrations directly on the 16 coordination points, simulating multi-motor vibrations.
To render spatial information of a static object using vibration, the touch probe first moves above of the center of the hand and then lowers to touch the skin. The vibration motor on the touch probe is triggered the moment contact is made. It then lifts up, move to the target point, lowers again, and vibrates upon contact (Figure \ref{fig:vib_skin_move}b). Each vibration lasts for \SI{0.5}{\second}, adhering to the common standard of default vibration length used in commercial devices such as smartphones or smartwatches.
The rendering of the moving trajectory follows the same principle. The touch probe lowers at each point along the mapped path and renders vibration accordingly. The overall haptic rendering time remains consistent with that of the skin-stretch condition.

\subsection{Task One: Understanding Spatial Information of a Static Object}
In Task One, we investigated how blind individuals perceive static spatial information (e.g., the direction and distance of static virtual objects) through skin-stretch and vibrotactile haptic feedback.

\subsubsection{Task procedure}
Task One required approximately \SI{60}{\minute} to complete. The study was designed as a within-subjects experiment, with the order of experiencing two types of haptic feedback counterbalanced.

\paragraph{Introduction and learning} 
We first introduced the haptic device to the participants, explaining how it renders both vibrotactile and skin-stretch feedback on the dorsal side of the hand.
We then instructed participants to place their non-dominant hand on the resting pad. 
We chose to render the haptic cues on the non-dominant hand for two reasons: first, the non-dominant hand may posses more precise haptic perceptions than dominant hand~\cite{squeri2012two}; second, adding haptic feedback to the dominant hand could overwhelm the users in cases when their dominant hand is being used to interact with other VR content.
After the participant placed their hand on the resting pad, we conducted the calibration procedure and then asked them to experiment with both types of haptic feedback at various points on their dorsal side of the hand. We explained how this haptic rendering could be interpreted as the spatial information of a virtual object. 
We then began the testing phase.

\paragraph{Testing} 

As described in Section~\ref{haptic pattern design}, there were 16 different points on the dorsal side of the hand, representing 8 directions and 2 levels of distance. 
We tested participants' perceptions at each point for both haptic conditions.
Each point was tested three times, resulting in a total of 48 trials for each type of haptic feedback.
The order of the trials was randomized. 
After completing each trial, participants were asked to describe the distance (i.e., near or far) and direction they perceived. 
We recorded the responses and calculated the accuracy for further analysis. To minimize any carryover effects, correctness feedback was not provided to participants after each trial.
Participants were given a \SI{5}{\minute} break between the two types of haptic feedback trials, with additional breaks provided whenever requested.

\paragraph{Exiting interview and questionnaire}
After completing one type of haptic feedback experiment (e.g., all 48 trials for skin-stretch), participants were asked to complete the NASA-TLX questionnaire~\cite{nasa_tlx} to assess task load. 
Additionally, Likert-scale questions were posed to gauge participants' confidence in their responses. 
Upon finishing the entire test, a semi-structured interview was conducted, during which participants were asked to compare the two types of haptic feedback.

\begin{figure}[h!]
    \includegraphics[width=\columnwidth]{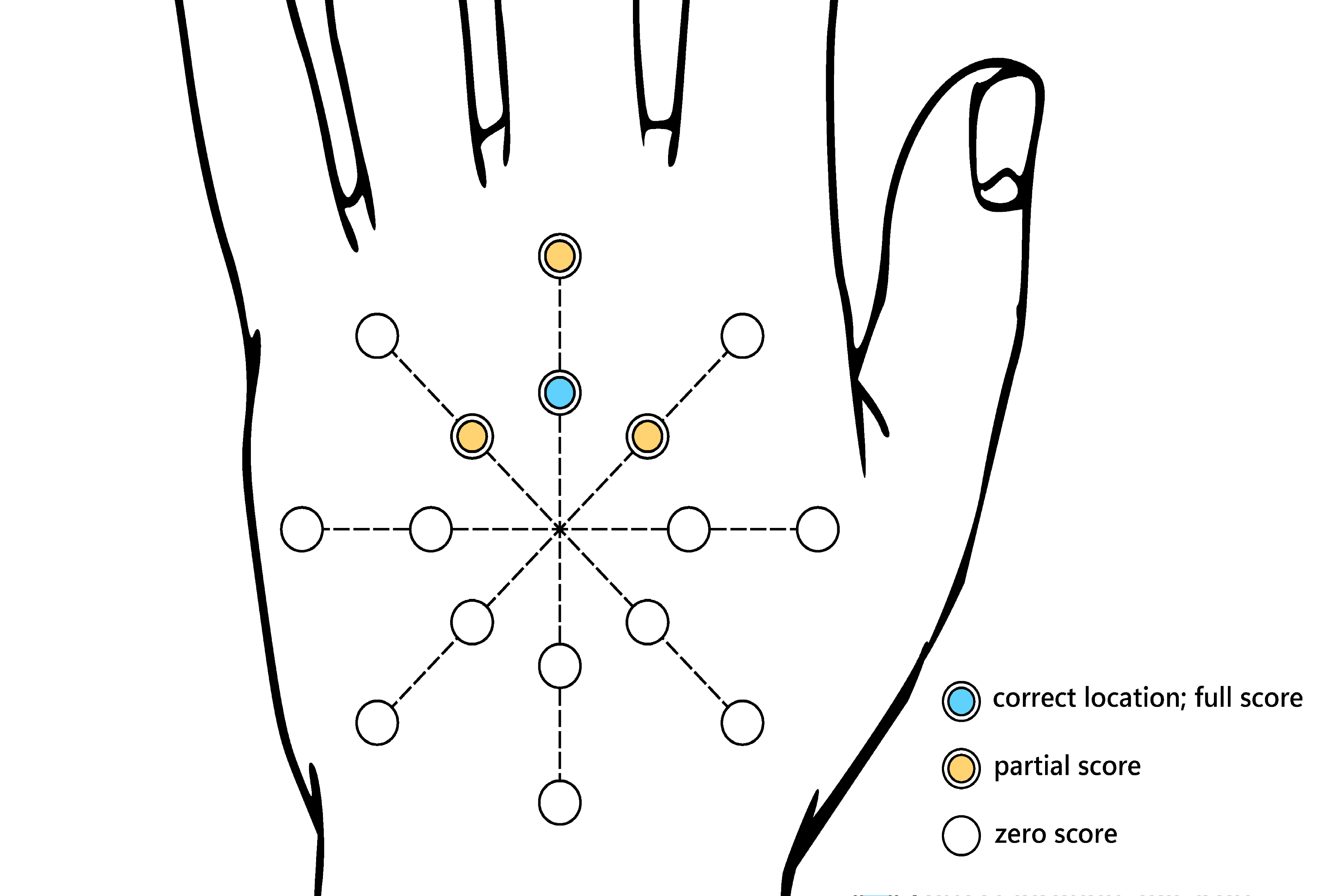}
    \caption{Score criteria: A correct answer, marked in blue, receives full points. A partially correct answer, marked in yellow, receives half points. An incorrect answer, marked in white, receives no points.}
    \label{fig:scoring}
\end{figure}

\subsubsection{Data collection and analysis}
We collected responses from a total of 960 trials, involving 10 participants each completing 48 trials for each type of haptic feedback. 
To evaluate participants' perception of spatial information at each coordinate point, we developed a scoring system. The scoring system uses 1 point for a full score, 0.5 points for a partial score, and 0 for no score (Figure~\ref{fig:scoring}). 
A full score (1 point) was awarded only if the participant accurately identified both distance and directional information (e.g., Figure \ref{fig:scoring}, the point marked in blue is the correct position).
A partial score (0.5 points) was given if the participant's response was within the nearest points to the correct location (e.g., Figure \ref{fig:scoring}, points marked in yellow).
A zero score (0 points) was assigned for responses outside the correct or partial correct range (e.g., Figure \ref{fig:scoring}, points marked in white).

\begin{figure}[t]
    \label{fig:task1_data_differences}
    \includegraphics[width=\columnwidth]{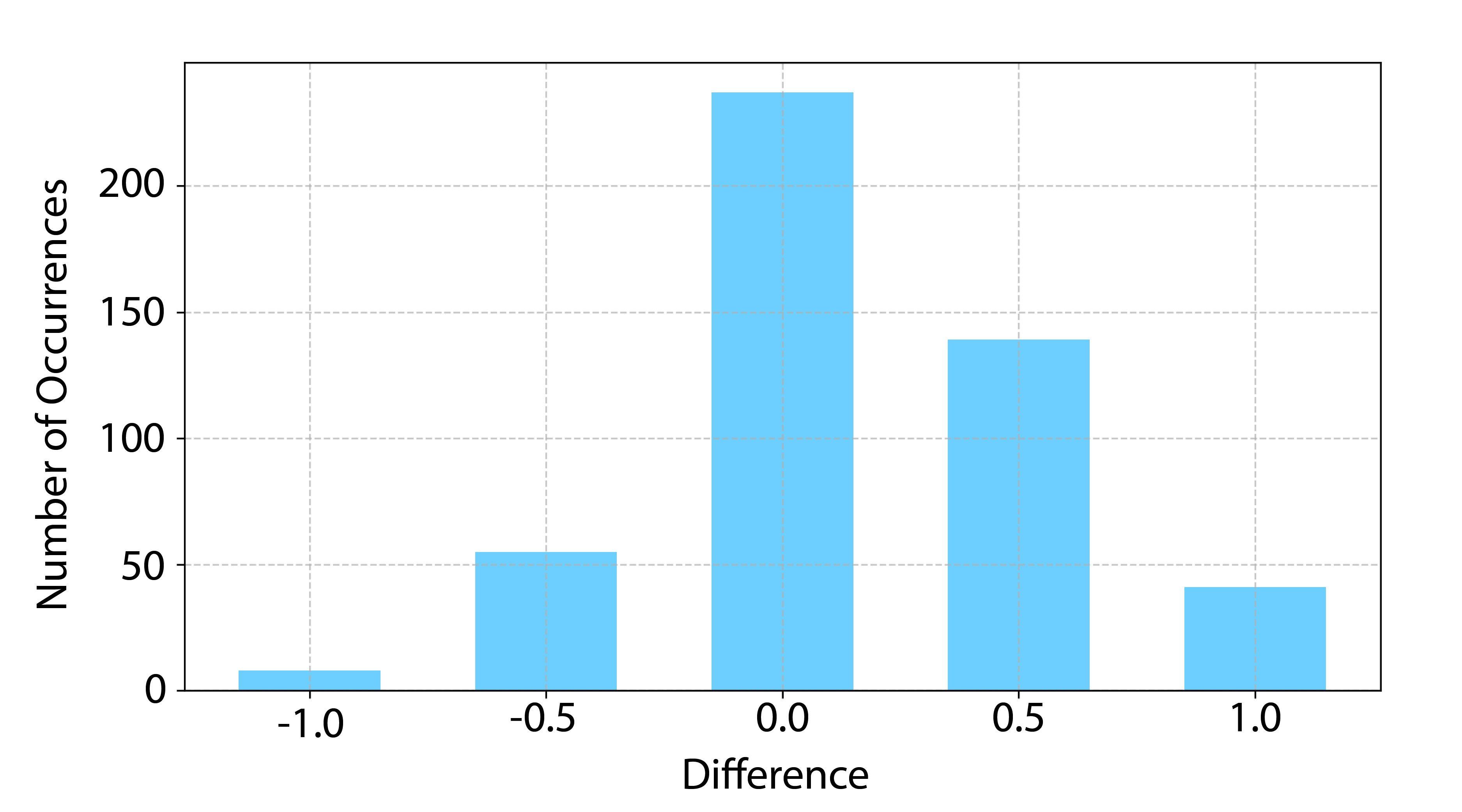}
    \caption{The distribution of paired-score differences in Task One. The differences were approximately symmetrically distributed around zero.}
    \label{fig:task1_diff}
\end{figure}

Our study included a within-subject factor, \textit{haptic mechanism}, with two levels: \textit{vibration} and \textit{skin-stretch}. 
Perception accuracy for each point was calculated as the average score across three trials.
The Shapiro-Wilk test indicated that the scores did not follow a normal distribution ({\slshape W = 0.76, p < .001}). Additionally, the difference between the scores were approximately symmetrically distributed around zero (Figure \ref{fig:task1_data_differences}). We therefore used a non-parametric Wilcoxon signed-rank test to compare participants' performance across the two conditions.

Similarly, as the responses from the NASA-TLX and the Likert-scale scores did not follow a normal distribution, we applied the Wilcoxon signed-rank test for non-parametric analysis.

\subsection{Task Two: Understanding Spatial Information of a Moving Object}

In Task Two, we investigated how blind participants perceived the spatial information of a moving object. 

\begin{figure}[h]
    \includegraphics[width=\columnwidth]{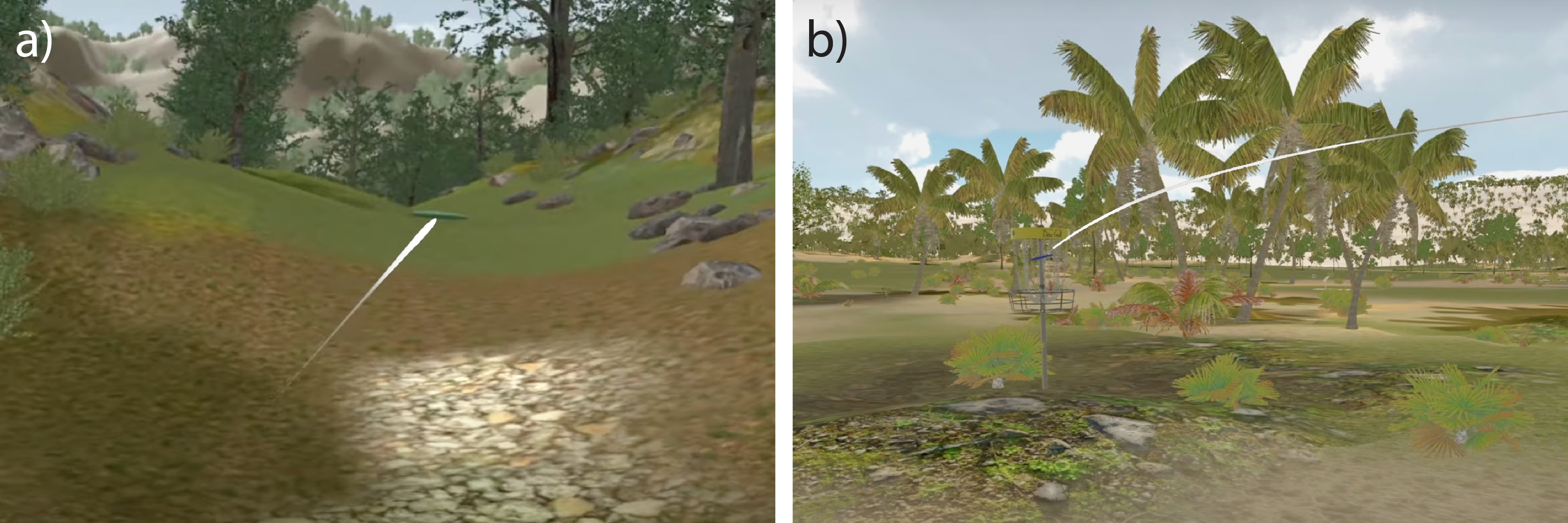}
  \caption{Disc trajectories in a disc golf VR game: a) a single straight line, and b) a parabola curve.}
  \label{fig:gameshot}
\end{figure}

\begin{figure}[h!]
    \includegraphics[width=\columnwidth]{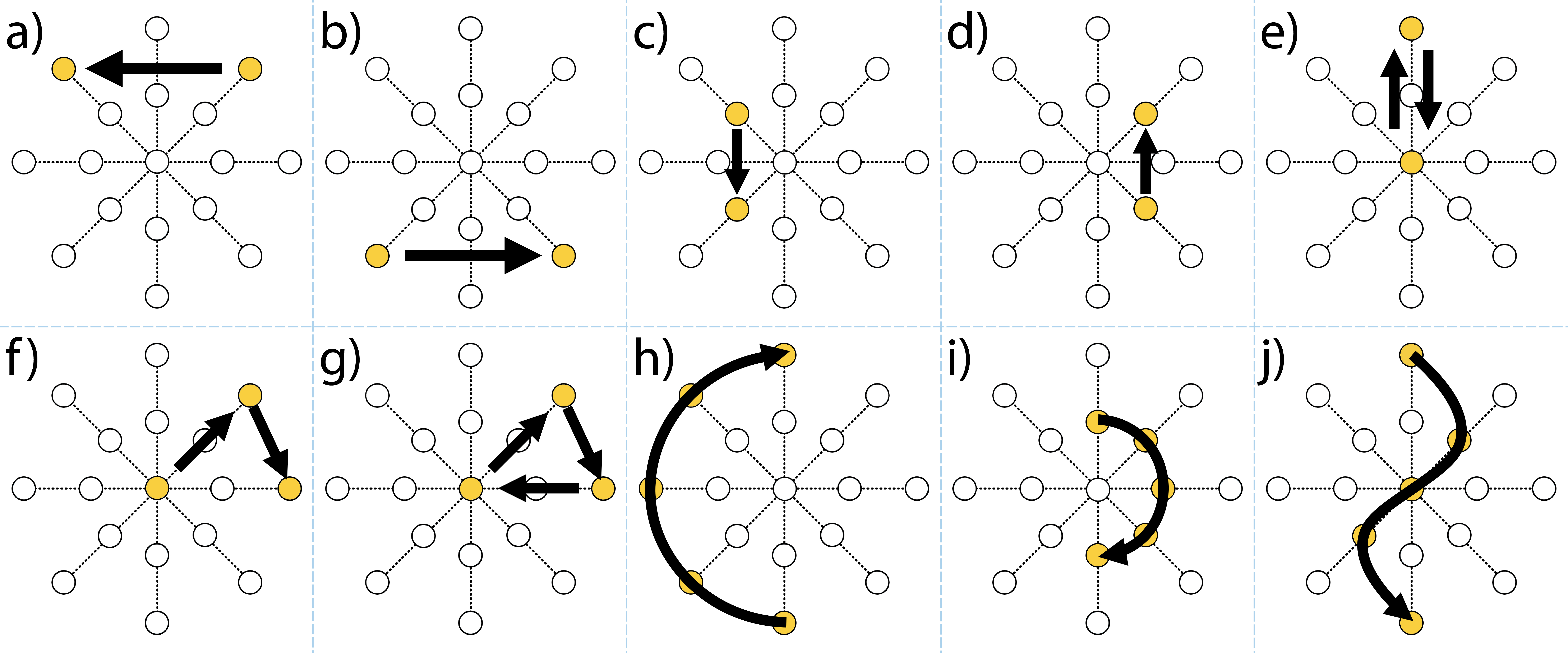}
    \caption{Ten trajectories for rendering the spatial information of a moving object.}
    \label{fig:dynamic_patterns}
\end{figure}

As the spatial information of a moving object constitutes a collection of discrete spatial information over a period of time, it is impractical for participants to simply describe it in detail (e.g., the touch probe starts moving at the point in the front direction at far distance, then passing the center point of the hand, and finally stops at the far-back distance point). 
Thus, to help participants better understand the moving  trajectory in virtual space, we referenced a golf disc game (Figure~\ref{fig:gameshot}) in VR. 
Participants were told that the touch probe represented the position of the flying disc in the virtual space. 
Specifically, we introduced ten test paths (Figure~\ref{fig:dynamic_patterns}), categorized into three types: (1) single-direction moving trajectories (2) multiple-direction polygon trajectories, and (3) curved and S-shaped trajectories. 
The single-direction paths simulated scenarios where the user throws the disc in a straight path or other players throw the disc toward the user. 
The multiple-direction polygon trajectories simulated a game where two or more players throw the disc between each other. 
The curve and S-shaped paths simulated a flying disc changing directions.

We rendered both types of haptic feedback across multiple trials. Participants were asked to verbally identify the starting and ending points of each path. 
In addition, participants were asked to select the moving path from one of three options: a straight-line path, multiple-direction path, or a curved or S-shaped path. 
Participants were also allowed to use their dominant hand to reproduce the touch probe's movement if they preferred.
The test trials began once participants felt comfortable with the haptic rendering. All trials were counterbalanced.

\subsubsection{Task procedure}
Task Two lasted approximately \SI{60}{\minute}. 
As it used the same haptic device and rendered the same types of haptic feedback, the procedure for Task Two closely resembled that of Task One.
Participants completed one trial for each moving path, resulting in a total of 20 trials across both skin-stretch and vibrotactile haptic feedback.

\subsubsection{Data collection and analysis}
We collected a total of 200 trials, calculated as 10 participants $\times$ 10 trials $\times$ 2 haptic feedback types (skin-stretch and vibration).
Each valid response from a trial included three components: the moving object's starting point, ending point, and type of moving path.
All these components were equally important, as misunderstanding any of them could confuse users and potentially diminish their experience in VR games. 

We employed a binary coding system to evaluate all three components. A correct response for each part of a trial was recorded as one, while an incorrect response was recorded as zero.
Thus, the score of each trial could range from 0 to 3.
We calculated the accuracy of each trial by computing the ratio of the each trial's score to the maximum possible score of three, as shown below:
\[\frac{starting\; point + ending\; point + path}3\] 

We analyzed the accuracy of participants' response in conveying the motion path for the two types of haptic feedback.
The Shapiro-Wilk test indicated that the data did not meet the normality assumption ({\slshape W = 0.82, p < .001}).

Therefore, we proceeded with non-parametric analyses using the Wilcoxon signed-rank test.
As in Task One, we also collected responses from the NASA-TLX and Likert-scale questions and analyzed them using the Wilcoxon signed-rank test.

\section{Result}\label{result}

\subsection{Task One: Spatial Information of a Static Object}

\begin{figure}[t]
\centering
    \includegraphics[width=\columnwidth]{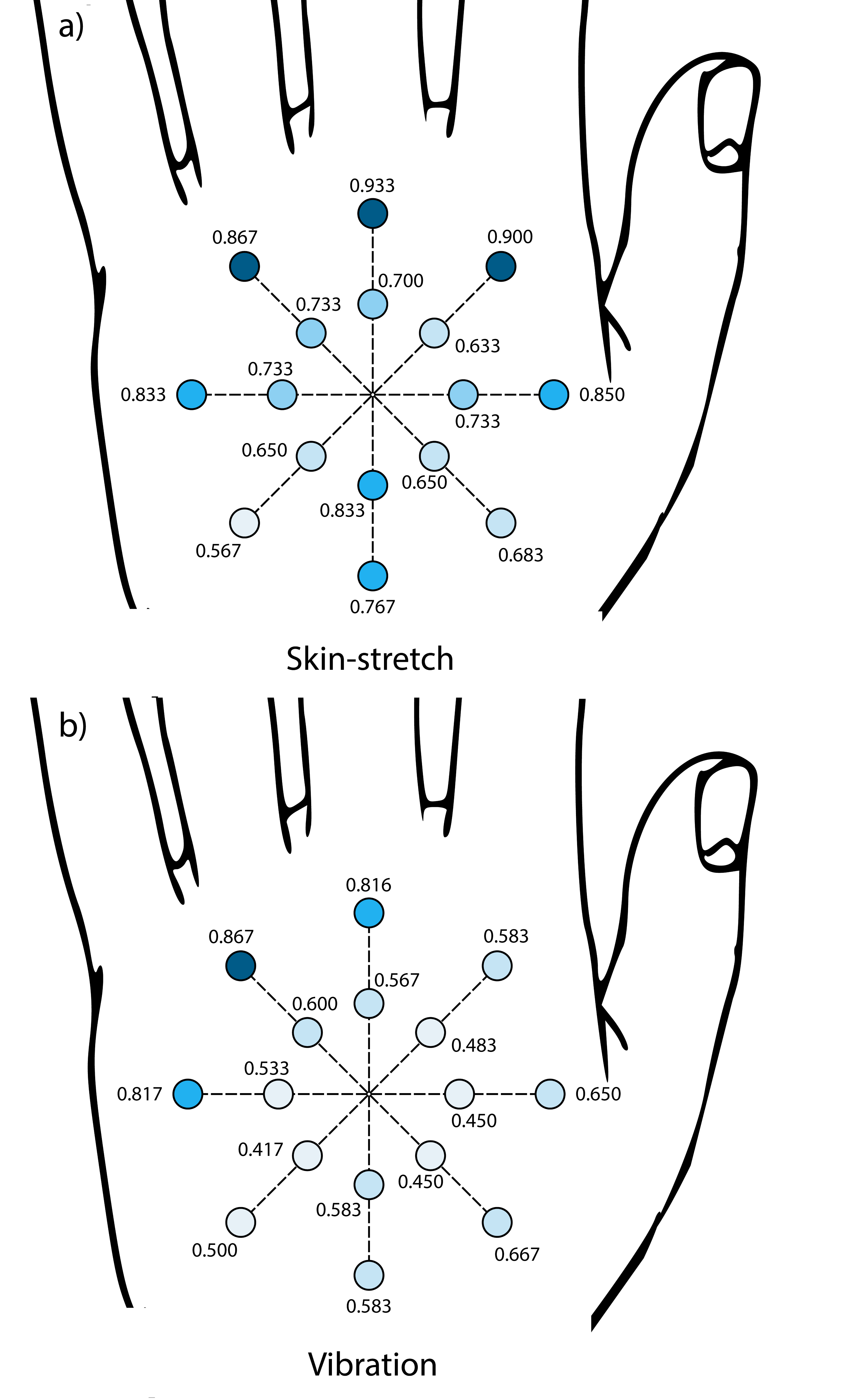}
    \caption{Task One: Scores at each point under a) skin-stretch mode and b) vibration mode.}
    
    \label{fig:task1_result}
\end{figure}

\subsubsection{Direction and distance}
We calculated the mean scores from 960 trials across the two haptic feedback types, with higher scores indicating greater accuracy.
The paired Wilcoxon signed-rank test revealed a significant effect of haptic mechanisms on participants' accuracy in perceiving spatial information ({\slshape Z = 7.64, p < .001, r = 0.25}). 
Participants showed higher accuracy with skin-stretch feedback ({\slshape M} = 0.754, {\slshape SD} = 0.313) compared to vibration feedback ({\slshape M} = 0.598, {\slshape SD} = 0.371).

To explore spatial perception across different areas of the back of the hand, we analyzed the scores for each point on the dorsal side for both feedback types (Figure \ref{fig:task1_result}a and b). 
For both skin-stretch and vibration, the upper area of the hand was more sensitive and yielded higher accuracy compared to the lower area.
This may be because that these regions are near distal joints, which provide clear reference points for haptic perception~\cite{collins2005cutaneous, edin2001cutaneous, bark2008comparison}.
For example, participants accurately perceived feedback near the middle finger's knuckle when the touch probe rendered far-front locations.
Conversely, lower scores in the lower regions of the hand may be due to the lack of distinct reference points, such as knuckles or the wrist joint, making it more challenging for participants to determine direction and distance.

\begin{figure}
    \includegraphics[width=\columnwidth]{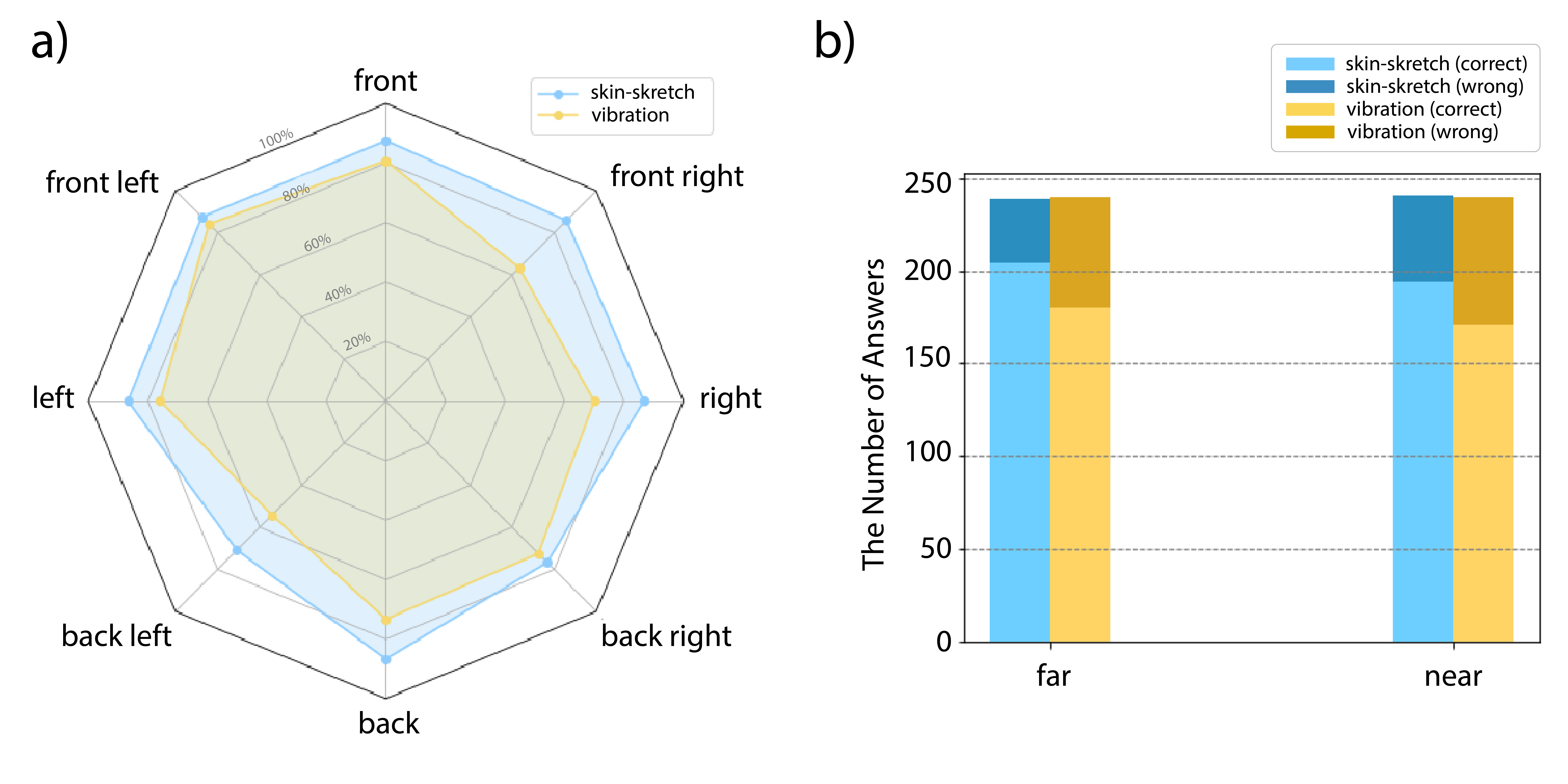}
    \caption{Task One: a) The accuracy of eight directions. b) The accuracy in the two distances.}
    \label{fig:task1_dir_dis}
\end{figure}

We also examined directional perception. 
Scores for each direction were calculated based on correct directional responses, regardless of distance accuracy (Figure~\ref{fig:task1_dir_dis}a). 
Skin-stretch feedback outperformed vibration, with a mean score of 0.835 ({\slshape SD} = 0.259) compared to 0.718 ({\slshape SD} = 0.339).
Similarly, scores for near and far distances showed higher accuracy for skin-stretch feedback (far: {\slshape M} = 0.870, {\slshape SD} = 0.242, near: {\slshape M} = 0.801, {\slshape SD} = 0.270) than for vibration (far: {\slshape M} = 0.802, {\slshape SD} = 0.285, near: {\slshape M} = 0.633, {\slshape SD} = 0.367) (Figure~\ref{fig:task1_dir_dis}b).
A paired Wilcoxon test confirmed the significant effect of haptic feedback type on perception accuracy for both directions ({\slshape Z = 6.12, p < .001, r = 0.198}) and distances ({\slshape Z = 4.24, p < .001, r = 0.137}).

\subsubsection{NASA-TLX and Likert-scale questionnaire}

\begin{figure}[h]
    \includegraphics[width=\columnwidth]{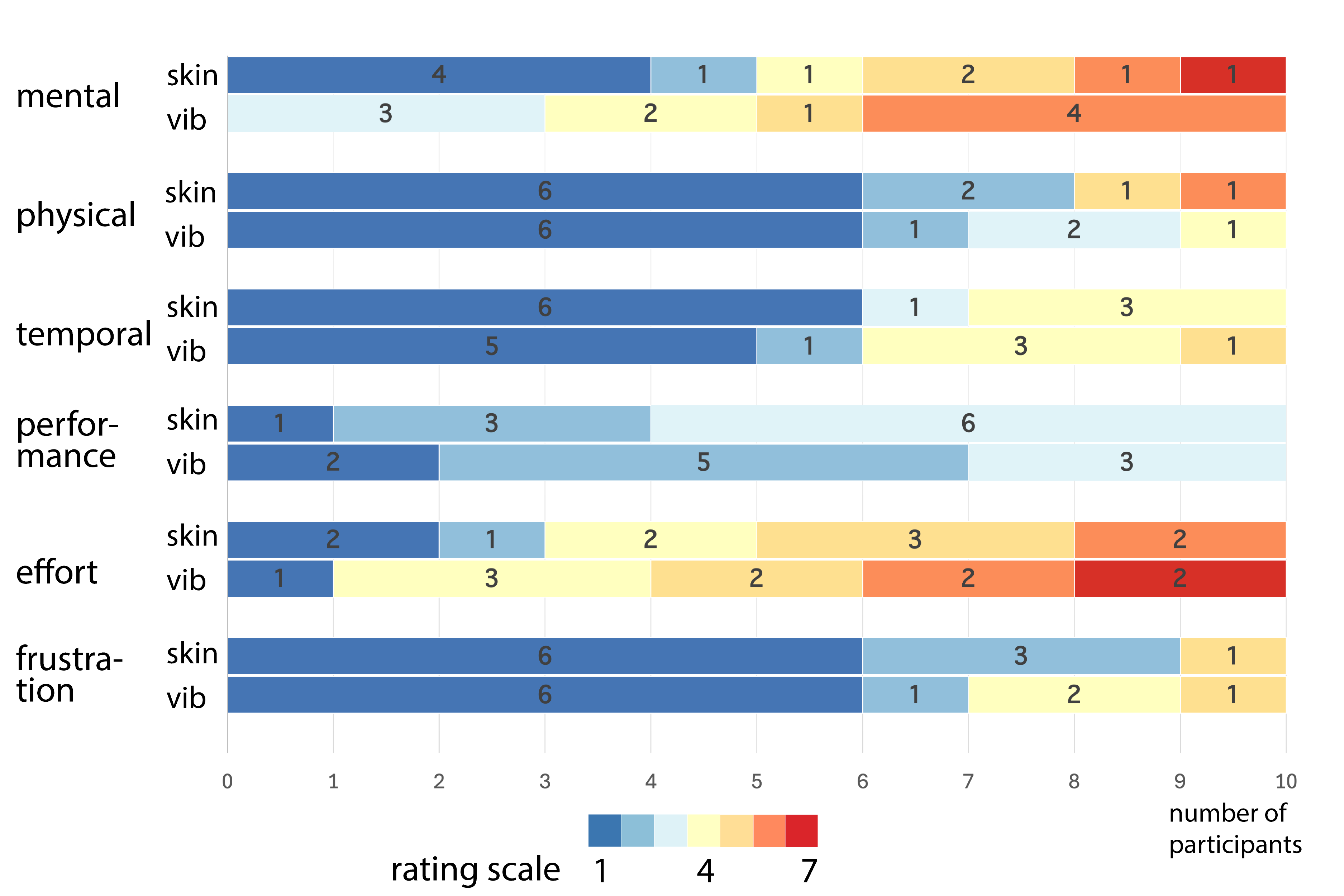}
    \caption{
Self-reported NASA-TLX ratings for Task One, with a scale ranging from 1 (indicating the lowest perceived workload, coded in red) to 7 (indicating the highest perceived workload, coded in blue).}
    \label{fig:task1_nasa}
\end{figure}

The task load ratings for each haptic feedback mechanism, as measured by the NASA-TLX, are shown in Figure~\ref{fig:task1_nasa}.
Specifically, the task load ratings for the skin-stretch mechanism ({\slshape M} = 2.6, {\slshape SD} = 0.83) was lower than those for vibrotactile feedback ({\slshape M} = 3.0, {\slshape SD}  = 1.38).
Notably, skin-stretch imposed significantly lower mental load on participants (3.3 vs 4.6) compared to vibration.
However, no significant effect of haptic mechanisms was found on task load ratings ({\slshape Z = -1.76, p = 0.079, r = -0.16}).

\begin{figure}
    \includegraphics[width=\columnwidth]{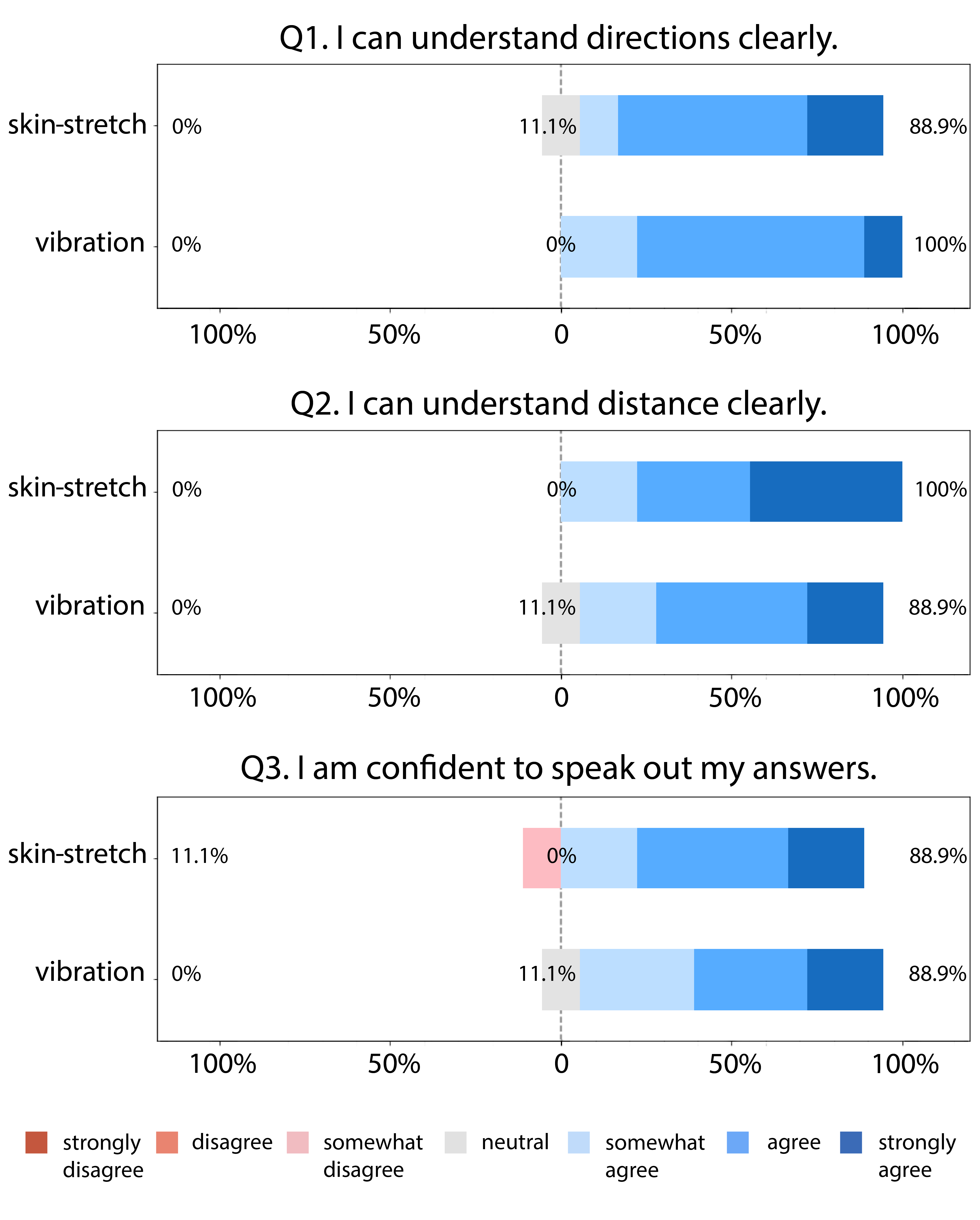}
    \caption{Self-reported ratings of spatial information understanding for Task One.}
    \label{fig:task1_likert}
\end{figure}

In addition to the NASA-TLX, we asked participants about their confidence in their answers using three Likert scale questions, with the results presented in Figure~\ref{fig:task1_likert}.
Participants reported that they could clearly understand distance information (skin-stretch: {\slshape M} = 6, {\slshape SD}  = 1.1; vibration: {\slshape M} = 5.8, {\slshape SD}  = 0.92) and directional information (skin-stretch: {\slshape M} = 5.8, {\slshape SD}  = 0.92; vibration {\slshape M} = 5.9, {\slshape SD}  = 0.57) from both haptic feedback types. They also expressed confidence in providing their answers (skin-stretch: {\slshape M} = 5.5, {\slshape SD}  = 1.27; vibration: {\slshape M} = 5.7, {\slshape SD}  = 0.95).

However, we did not find a significant effect of haptic feedback mechanisms on participants' understanding of distance or directional information, or on their confidence (distance: {\slshape Z = 0.27, p = 1.0, r = 0.06}; direction: {\slshape Z = 0.42, p = 0.84, r = 0.09}; confidence: {\slshape Z = 0.05, p = 1. r = 0.01}).

\subsubsection{Post-task interview}

Interestingly, although skin-stretch mechanism showed an overall higher accuracy in rendering spatial information, preferences among participants were divided.
Several participants favored vibration cues, citing greater familiarity with vibration-based feedback in their daily experiences.
\begin{quote}
{\itshape "For me, the vibration (haptic feedback) is much easier to tell the direction and near or far (distance on my back of the hand) than the skin-stretch (haptic feedback). I am more familiar with the vibration (haptic feedback), because my phone can vibrate. I can get the notification sounds, but it also vibrates (to let me know)."} ---P4
\end{quote}

\begin{quote}    
{\itshape "...I didn't get (all directions), but I think the vibration is a little easier to judge the angles (than the skin-stretch)." } --- P2
\end{quote}

Other participants stated that the skin-stretch haptic feedback was more practical and useful for them.
\begin{quote}
{\itshape 
"I like the skin-stretch (haptic feedback)... because the rubber tip will pull my skin while it is moving, so it can give me a clear feeling. The vibration will always start at the center point on my back of hand, which is good, but the problem is ... I need to pay attention to figure out where is (the vibration), and I need to anticipate it.... It's sort a mental thing, because for the skin-stretch, I have already knew (the touch tip) is pulling me, and I don't need to anticipate where the (touch tip) going... I think the skin-stretch is more nature to me." ---P1}
\end{quote}

\begin{quote}
{\itshape 
"...For the skin-stretch, I can feel the pull and stops exactly at near or far, so I think it may be easier for me to know the distance, but for vibration I am just guessing it is near or far....". ---P6
}
\end{quote}

Additionally, some participants suggested that a combination of both mechanisms might provide the best experience.

\begin{quote}
{\itshape 
"...I think combining both vibration and skin-stretch would be better. For example, using skin-stretch to know the directions and using the vibrations to know the distances." ---P5
}
\end{quote}

\subsection{Task Two: Spatial Information of a Moving Object}

\subsubsection{Accuracy in perceiving a moving object's path.}
\begin{table*}[]
\centering
\setlength\tabcolsep{4pt} 
\caption{Average perceptual accuracy for different moving patterns in Task Two }
\begin{tabular}{llllllllllll}
\hline
Patterns &
\begin{minipage}{0.06\textwidth}
    \includegraphics[width=1cm]{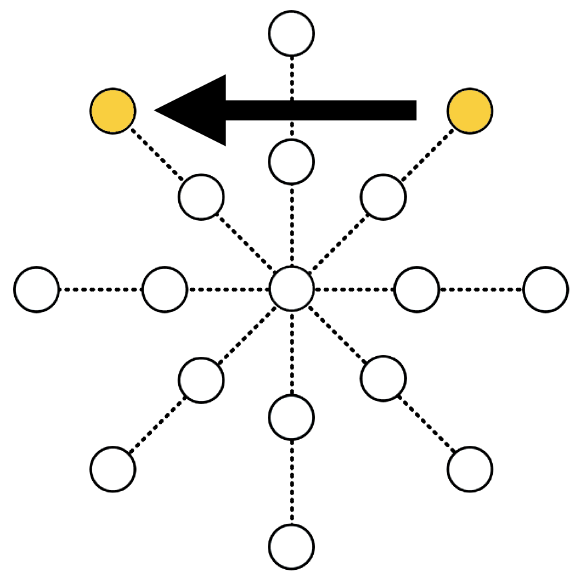}
\end{minipage} & 
\begin{minipage}{0.06\textwidth}
    \includegraphics[width=1cm]{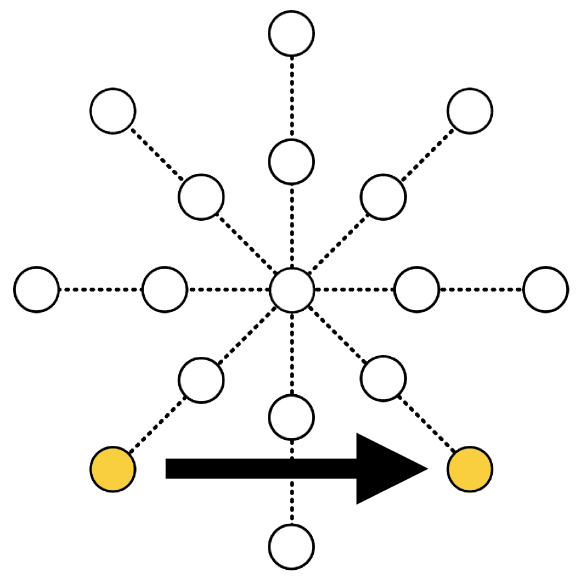}
\end{minipage} & 
\begin{minipage}{0.06\textwidth}
    \includegraphics[width=1cm]{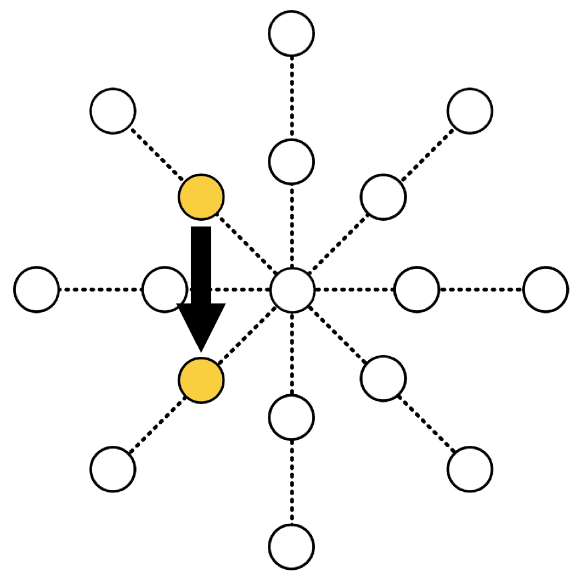}
\end{minipage} & 
\begin{minipage}{0.06\textwidth}
    \includegraphics[width=1cm]{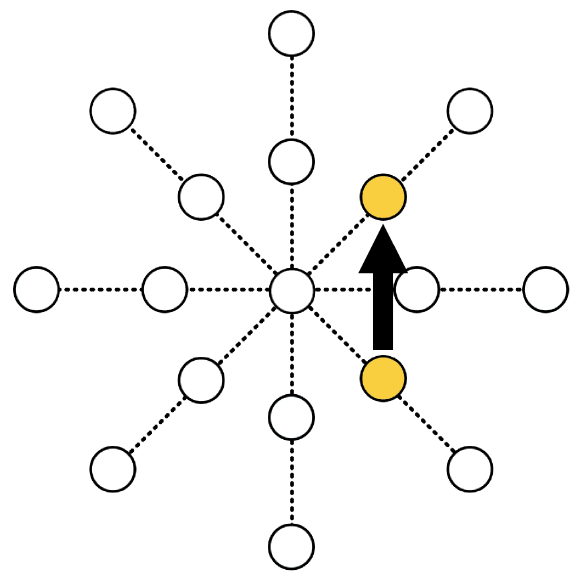}
\end{minipage} & 
\begin{minipage}{0.06\textwidth}
    \includegraphics[width=1cm]{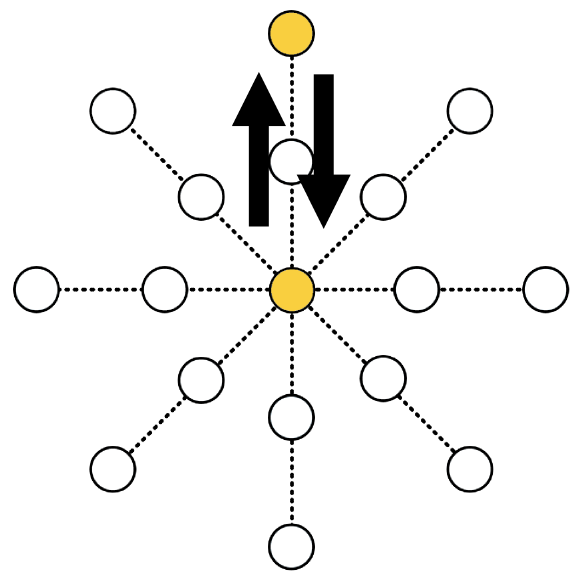}
\end{minipage} & 
\begin{minipage}{0.06\textwidth}
    \includegraphics[width=1cm]{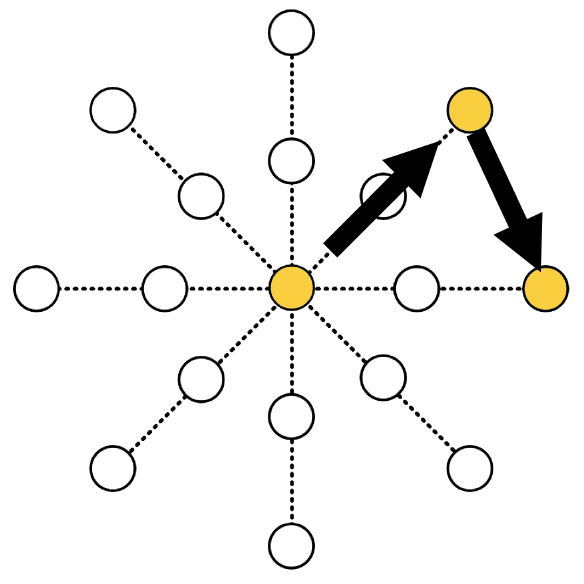}
\end{minipage} & 
\begin{minipage}{0.06\textwidth}
    \includegraphics[width=1cm]{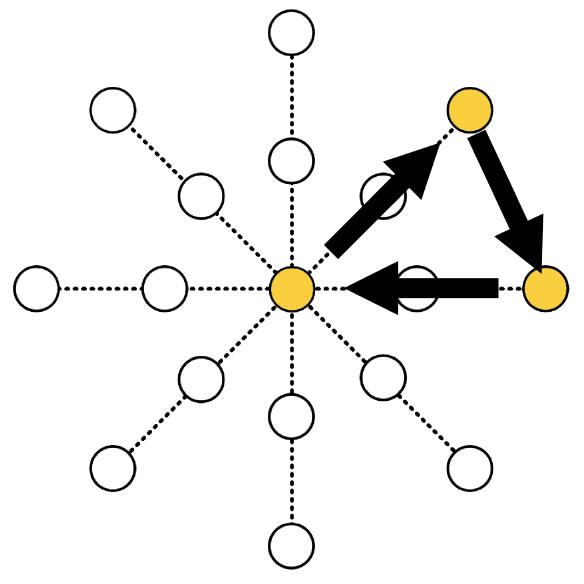}
\end{minipage} & 
\begin{minipage}{0.06\textwidth}
    \includegraphics[width=1cm]{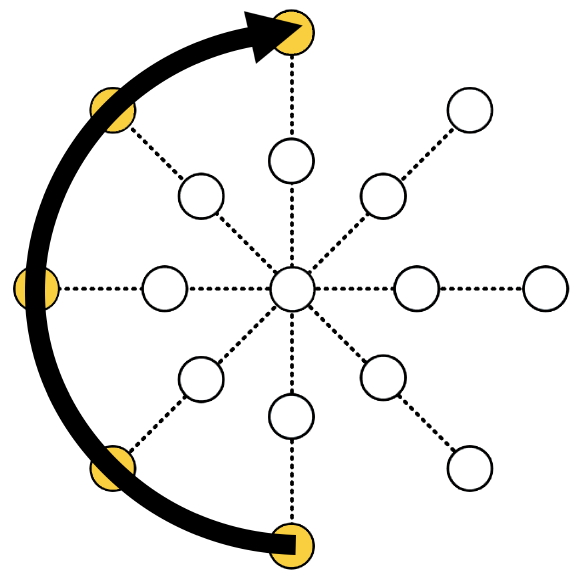}
\end{minipage} & 
\begin{minipage}{0.06\textwidth}
    \includegraphics[width=1cm]{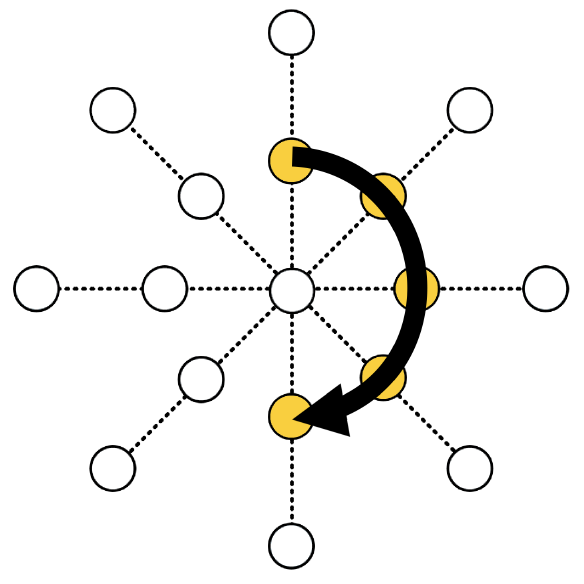}
\end{minipage} & 
\begin{minipage}{0.06\textwidth}
    \includegraphics[width=1cm]{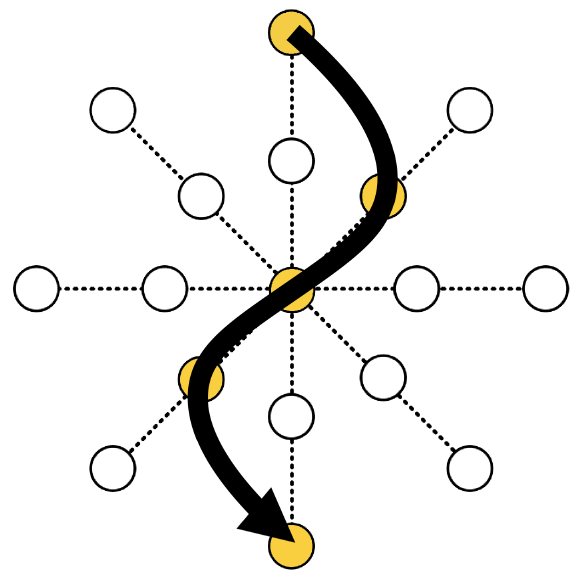}
\end{minipage} & 
Average \\ \hline

skin-stretch & 93.3\% & 93.3\% & 86.67\% & 73.3\% & 96.6\% & 70.0\% & 76.7\% & 66.7\% & 66.7\% & 56.7\% & 77.7\% \\
Vibration & 86.7\% & 76.7\% & 56.7\% & 60.0\% & 80.0\% & 70.0\% & 63.3\% & 46.7\% & 23.3\% & 43.3\% & 60.3\% \\ \hline
\end{tabular}

\label{tab:test2_pattern}
\end{table*}
We calculated accuracy over 200 trials between the two types of haptic feedback.
A paired Wilcoxon signed-rank test revealed a significant effect of haptic mechanisms on the accuracy with which blind participants perceived the moving object's spatial information ({\slshape Z = 4.7, p < .001, r = 0.33}).
Blind participants demonstrated higher accuracy with skin-stretch cues ({\slshape M} = 77.8\%, {\slshape SD} = 27.2\%) compared to vibrations ({\slshape M} = 60.6\%, {\slshape SD} = 32.3\%) in perceiving the moving object's spatial information.

In addition to evaluating overall accuracy, we examined how blind participants perceived different types of movement paths.
Table~\ref{tab:test2_pattern} presents the accuracy rates for all tested movement paths, including both skin-stretch and vibrotactile feedback. 
Accuracy was highest for single-direction movement paths, followed by multi-direction paths, with curved and S-shaped paths showing the lowest accuracy.
This pattern was consistent across both haptic feedback types.
Furthermore, skin-stretch feedback consistently resulted in higher (or equivalent) accuracy compared to vibrotactile haptic feedback across all path types.

\begin{figure}[h]
    \includegraphics[width=\columnwidth]{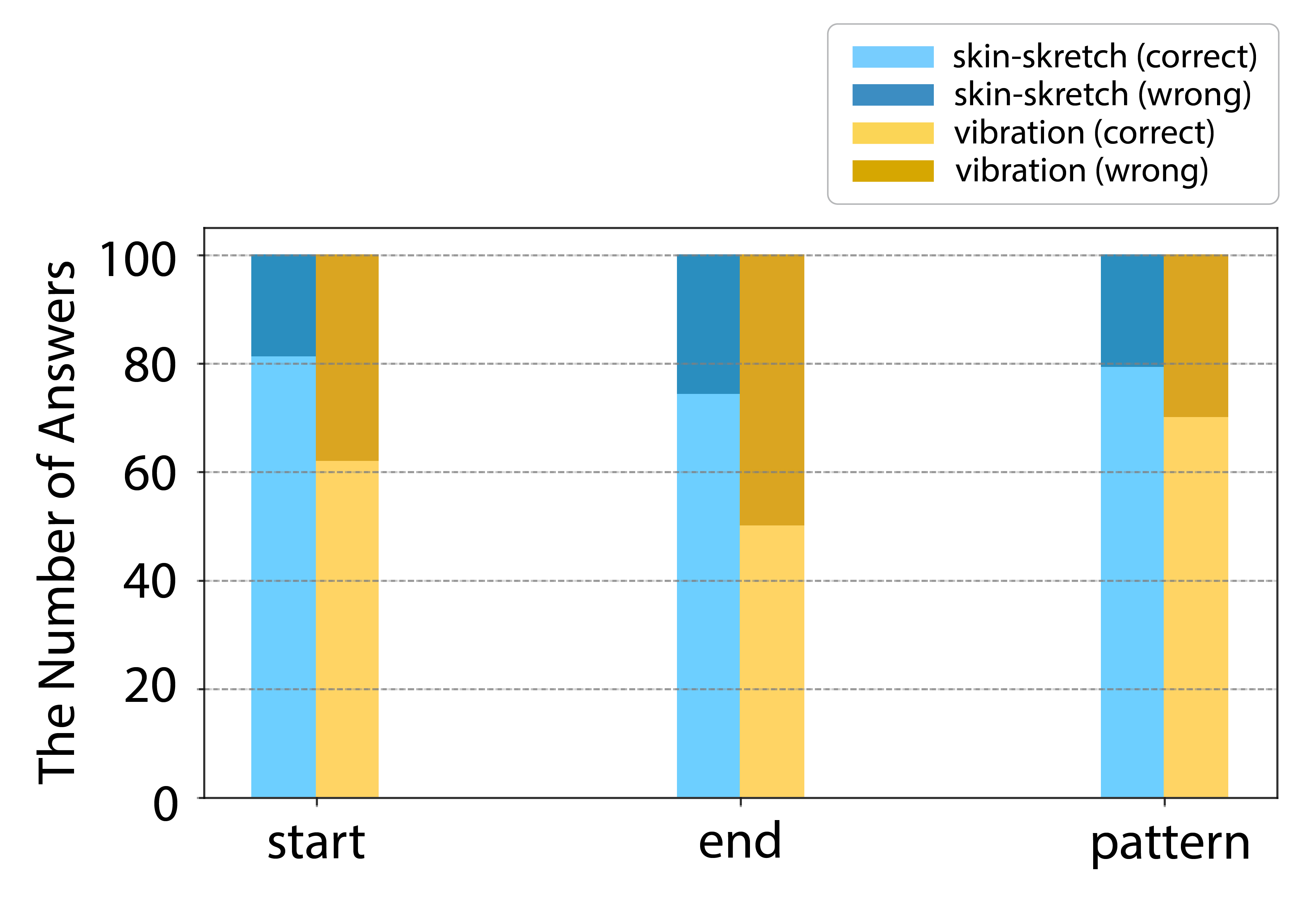}
    \caption{Average score for three components of Task Two.}
    \label{fig:test2_accuracy}
\end{figure}

To identify which components of the moving path (e.g., starting point, ending point, path pattern) posed challenges for blind participants, we further analyzed the accuracy for each components.
Figure~\ref{fig:test2_accuracy} presents the result.
Accuracy was lowest for identifying the ending point ({\slshape M}= 62.50\%) compared to the starting point ({\slshape M} = 71.50\%) and the movement pattern ({\slshape M} = 74.50\%).
One potential explanation is that participants might lose track of the reference point (e.g., the center point on the dorsum of the hand) during continuous haptic rendering, a challenge that may be more pronounced for longer moving paths, such as curved and S-shaped paths.

\subsubsection{NASA-TLX and Likert-scale questionnaire}
\begin{figure}[h!]
    \includegraphics[width=\columnwidth]{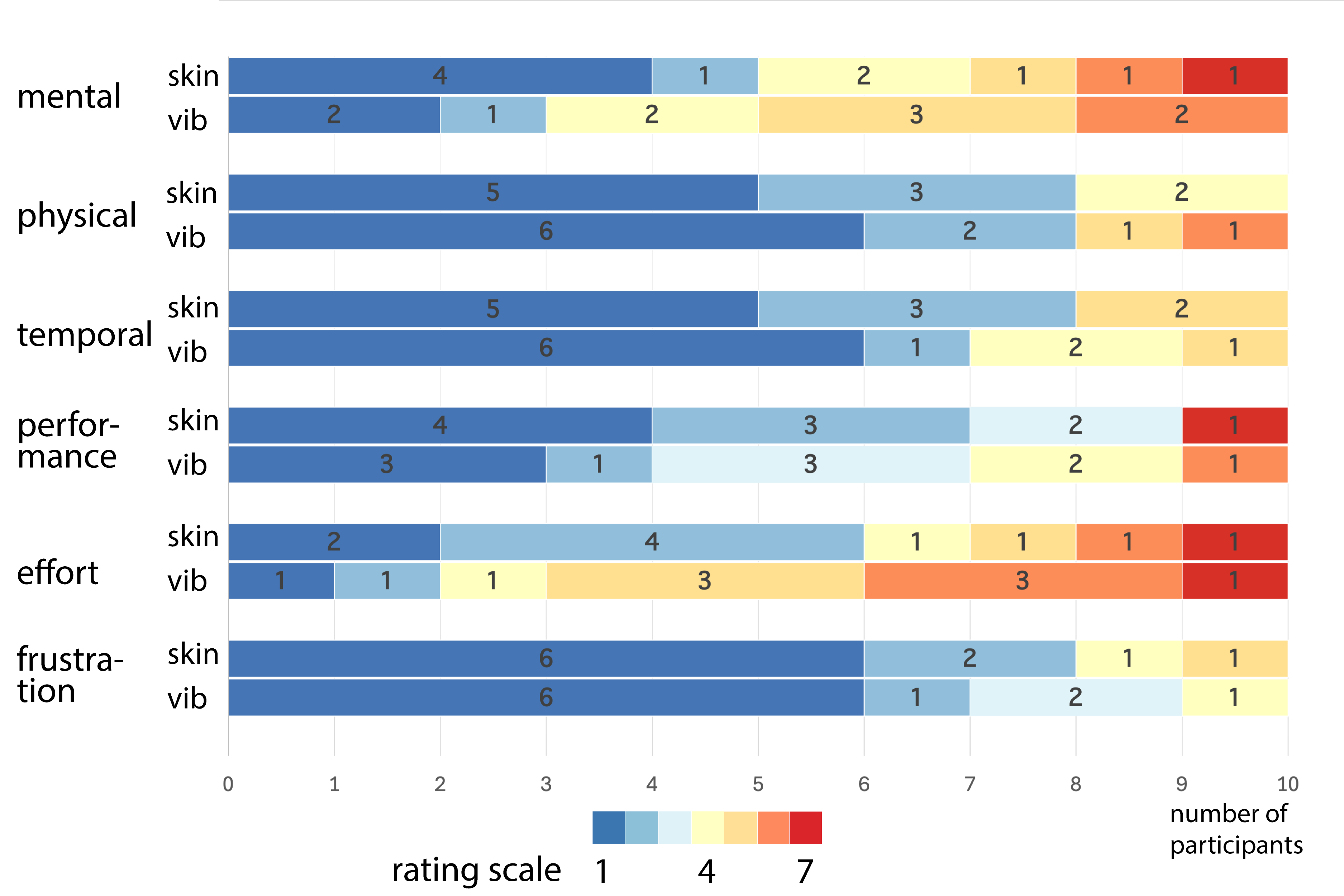}
    \caption{
Self-reported NASA-TLX ratings for Task Two, with a scale ranging from 1 (indicating the lowest perceived workload, coded in red) to 7 (indicating the highest perceived workload, coded in blue).}
    \label{fig:task2_nasa}
\end{figure}

The results of the NASA-TLX questionnaire are shown in Figure~\ref{fig:task2_nasa}.
The mental load rating for skin-stretch feedback ({\slshape M} = 2.4, {\slshape SD} = 0.61) was lower than that for vibration feedback ({\slshape M} = 2.9, {\slshape SD} = 1.16).
However, no significant effect of haptic mechanisms on task load ratings was found for blind participants perceiving the spatial information of moving objects ({\slshape Z = -1.57, p = 0.12, r = -0.14}).

\begin{figure}[h!]
    \includegraphics[width=\columnwidth]{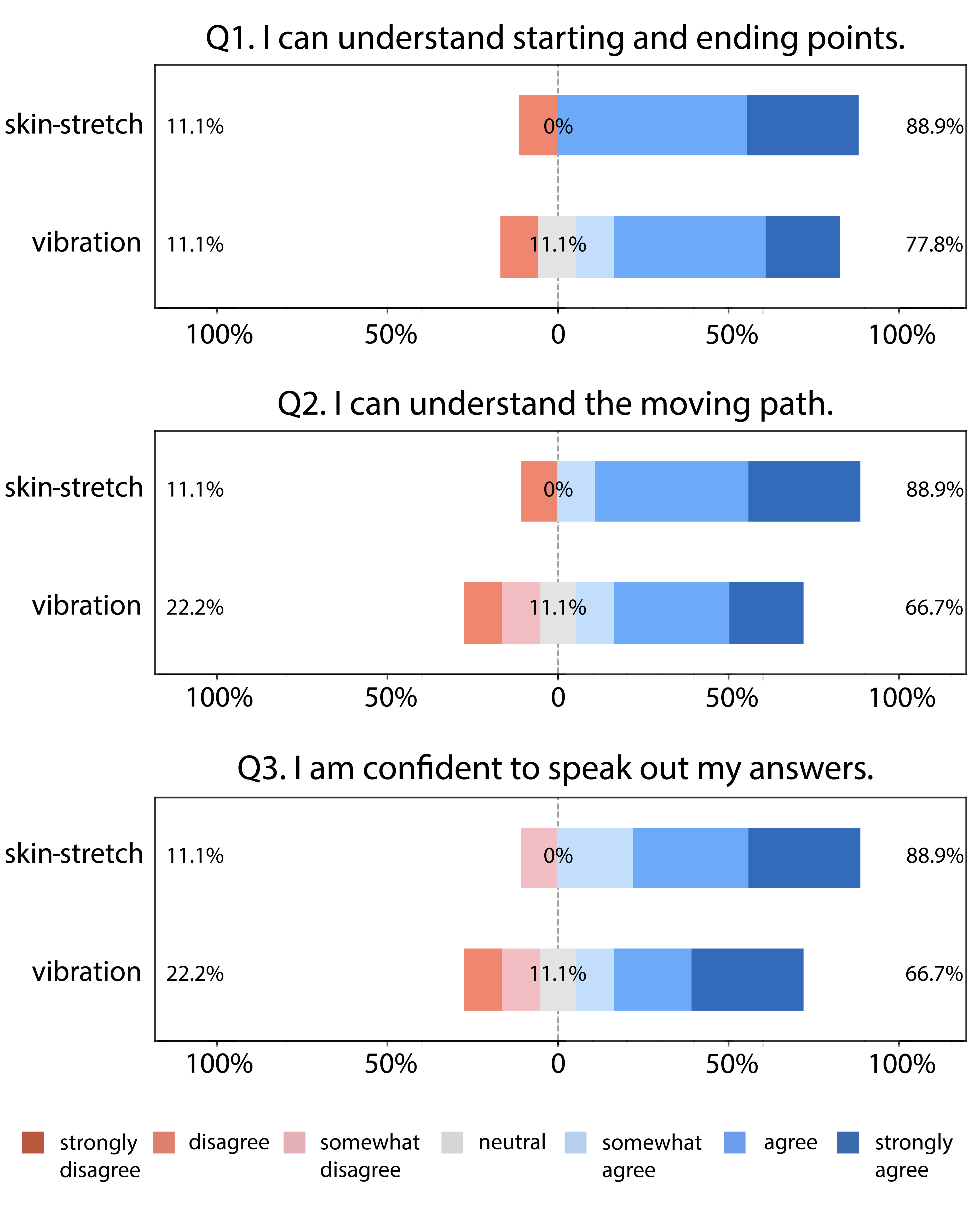}
    \caption{Self-reported ratings of understanding the moving object's spatial information in Task Two.}
    \label{fig:task2_likert}
\end{figure}

We also quantified participants' confidence in their responses using Likert scale questions (Figure~\ref{fig:task2_likert}).
Participants reported that they could understand the starting and ending points (skin-stretch: {\slshape M} = 5.9, {\slshape SD} = 1.5; vibration: {\slshape M} = 5.5, {\slshape SD} = 1.5), the object's moving path (skin-stretch: {\slshape M} = 5.8, {\slshape SD} = 1.5; vibration: {\slshape M} = 5.2, {\slshape SD} = 1.7), and that they could identify the overall moving path from the haptic cues rendered on the hand (skin-stretch: {\slshape M} = 5.9, {\slshape SD} = 1.3; vibration: {\slshape M} = 5.3, {\slshape SD} = 1.8).
However, no significant effect of haptic feedback mechanisms was found on participants' confidence in understanding the moving objects' starting and ending points ({\slshape Z = 1.1, p = 0.5, r = 0.1}), the distance ({\slshape Z = 1.0, p = 0.41, r = 0.09}), or their overall confidence in recognizing the paths ({\slshape Z = 1.3, p = 0.28, r = 0.12}).

\subsubsection{Post-task interview}

The ability to provide clear haptic feedback was key for participants to recognize the moving patterns. Several participants expressed a preference for skin-stretch, as it remained in constant contact with the skin as the probe moved, making it easier for users to perceive and identify patterns.
\begin{quote}
    {\itshape "...I prefer the skin-stretch because it is just showing the pattern. The vibration is fine, but it didn't show the patterns unless you are familiar with these patterns."---P2
    }
\end{quote}

Vibration, on the other hand, required greater focus and attention to discern locations on the back of the hand. However, a few participants, such as P7, preferred vibration over skin-stretch, as she found the vibrotactile sensation stronger and more familiar.

Additionally, we asked participants to share their thoughts on using haptic feedback and audio to provide spatial information about a moving object. 
They speculated the importance of synchronized audio and video in a virtual environment. 
Based on their daily experience with audio-based assistive tools, participants speculated that using audio to convey an object's location and movement would likely result in significant delays.
They also noted that relying on audio would likely require more effort, as they would need to constantly analyze the narration, remain attentive, and anticipate upcoming events.
In contrast, participants found the haptic feedback experienced during the study to be more responsive, providing a sensation akin to real-time feedback.

\begin{quote}
{\itshape 
"My brain is always running and I want something like this, which could synchronize my brain and feelings, because when I touch something I can direct feel it, this is one step. However, the audio description is two steps. You need first listen it, and then analyze it."---P1
}
\end{quote}

\section{Discussion and Limitation}

\subsection{Which Haptic Mechanism is Better?} %
Our study demonstrated that rendering haptics on the dorsal side of the user's hand can indeed help blind users perceive the spatial information of a static object and the trajectory of a moving object, although fully grasping the trajectory with precision remains challenging.

In terms of the two types of haptic mechanisms, the objective ratings suggested that skin-stretch outperformed vibration feedback in conveying spatial information for both static and moving objects. It is important to note that, although we used only one touch probe to simulate the multi-vibration condition, the 16 vibration points used represent one of the highest vibration densities observed in previous work. For example, Haptix~\cite{haptiX} explored the use of five vibration motors, and TactileGlove~\cite{TactileGlove} was equipped with up to nine vibrators. Given the size of the dorsal surface of a hand, it is unlikely that a haptic glove could provide more than 16 vibrations. Thus, this result indicates that continuous on-skin haptics will be better perceived by blind users compared to conveying the same information through discrete vibrations.

Interestingly, although the skin-stretch mechanism demonstrated higher accuracy, participants' subjective evaluations did not indicated a definitive preference for either of the two haptic mechanisms. In fact, as discussed in Section~\ref{result}, some participants preferred the tactile feedback from vibration over the skin-stretch mechanism, despite the latter's more accurate perception, citing greater familiarity. Unfortunately, the disparity in familiarity was not negligible in our study, as vibration feedback is indeed far more common than the skin-stretch mechanism. Although we provided a training session before each task, the differences in familiarity with the technologies were not completely overcome. It would be interesting to re-evaluate this preference in the future if both technologies can be equally well-known and accepted.

Nevertheless, we recognize the benefits of both haptic mechanisms for rendering spatial information. As some participants suggested, it should be feasible to combine them, e.g., by using skin-stretch to convey directional information and employing vibration to confirm the location once the touch probe reaches the target point on the skin. Future research should explore the potential of such a combination.

\subsection{Limitation}
Our study results should be interpreted with an understanding of the limitations regarding how broadly they can be applied.

For example, our study set the movement speed of the touch probe to \SI{8}{\milli\meter\per\second}. While this speed ensured participant safety, its capacity is limited in representing rapid changes in VR environments. How faster-moving objects or objects with varying speeds can be perceived haptically remains an open question.

Additionally, our experimental device is equipped with a single probe to deliver spatial information for one object at a time. 
Thus, the feasibility and efficacy of blind users perceiving multiple moving objects simultaneously through haptic feedback remains unclear.

Finally, in this current study, we used a desk-grounded device to generate haptic sensations. While this approach allowed us to test both mechanisms across large skin areas with one uniformed device, it also required participants to keep their hands in a fixed location. This condition does not reflect the dynamic manner in which users may interact with controllers in VR environments. Thus, although our study demonstrated promising results in providing spatial information through haptics to the dorsal hand, the findings may not fully extend to scenarios where individuals move their hands freely. Investigating how to accurately represent spatial information on a moving hand, as opposed to a stationary one, remains an open question and warrants further exploration in future research.

\subsection{Multi-Modal Feedback}
As the focus of our study is on haptic feedback, we intentionally omit audio in this work. In the future, we look forward to exploring multi-modal feedback, potentially integrating spatial audio with haptic feedback to provide more comprehensive information. While as we discussed with participants (P3), using audio can sometimes delay the rendering of information, the  audio can provide detailed information such as 3D spacing that is hard to achieve fully even with high-fidelity haptic solutions. Thus, it will be interesting to explore which types of spatial information, and in what contexts, should be rendered through spatial audio, haptics, or a combination of both.

\subsection{Minimization as a Wearable}
As discussed in the Introduction, our long-term vision is to develop compact, affordable, and effective haptic hardware that can be integrated into the existing VR hardware ecosystem, such as a hardware add-on to current VR controllers. Our current work represents one of the early steps toward this vision, with empirical results demonstrating that blind users can perceive spatial information via skin-stretch and vibrotactile haptics on the dorsal side of the hand. We anticipate that future work will build upon these empirical results to engineer wearable counterparts that can integrate with current VR systems and make VR truly accessible.

\section{Conclusion}
In this study, we investigated two types of haptic feedback—vibration and skin-stretch—for delivering spatial information about both static and moving virtual objects on the dorsal side of the hand for blind users. Our findings demonstrate that both mechanisms effectively convey spatial information, with the skin-stretch mechanism achieving higher accuracy. However, subjective evaluations from participants did not indicate a clear preference for either haptic mechanism. We discussed the strengths and limitations of both types of haptic cues and concluded with design recommendations for future haptic solutions aimed at improving VR accessibility.

\section*{Acknowledgment}
We thank all participants for their involvement in the user study. This work was supported in part by the National Science Foundation under Grant No. IIS-2328182 and a seed grant from the Maryland Initiative for Digital Accessibility. An LLM service was used solely for proofreading.

\bibliographystyle{abbrv-doi}

\appendix

\bibliography{my-reference}

\begin{thebibliography}{10}

\bibitem{3dSystemsPhantomPremium}
3d systems phantom premium.
\newblock \url{https://www.3dsystems.com/haptics-devices/3d-systems-phantom-premium}.
\newblock Accessed: Sept. 16, 2024.

\bibitem{play_blind}
R.~Andrade, M.~J. Rogerson, J.~Waycott, S.~Baker, and F.~Vetere.
\newblock Playing blind: Revealing the world of gamers with visual impairment.
\newblock In {\em Proceedings of the 2019 CHI Conference on Human Factors in Computing Systems}, CHI '19, p. 1–14. Association for Computing Machinery, New York, NY, USA, 2019. doi: {{%
10\hspace{.1pt}\discretionary{.}{%
}{.}\hspace{.4pt}1145\discretionary{/}{%
}{/}3290605\hspace{.1pt}\discretionary{.}{%
}{.}\hspace{.4pt}3300346}}


\bibitem{azenkot2024xr}
S.~Azenkot.
\newblock Xr access: Making virtual and augmented reality accessible to people with disabilities.
\newblock In {\em 2024 IEEE Conference Virtual Reality and 3D User Interfaces (VR)}, pp. xxvi--xxvi. IEEE, 2024.

\bibitem{bardot}
S.~Bardot, S.~Rempel, B.~Rey, A.~Neshati, Y.~Sakamoto, C.~Menon, and P.~Irani.
\newblock Eyes-free graph legibility: Using skin-dragging to provide a tactile graph visualization on the arm.
\newblock In {\em Proceedings of the 11th Augmented Human International Conference}, AH '20. Association for Computing Machinery, New York, NY, USA, 2020. doi: {{%
10\hspace{.1pt}\discretionary{.}{%
}{.}\hspace{.4pt}1145\discretionary{/}{%
}{/}3396339\hspace{.1pt}\discretionary{.}{%
}{.}\hspace{.4pt}3396344}}


\bibitem{bark2010rotational}
K.~Bark, J.~Wheeler, P.~Shull, J.~Savall, and M.~Cutkosky.
\newblock Rotational skin stretch feedback: A wearable haptic display for motion.
\newblock {\em IEEE Transactions on Haptics}, 3(3):166--176, 2010.

\bibitem{bark2008comparison}
K.~Bark, J.~W. Wheeler, S.~Premakumar, and M.~R. Cutkosky.
\newblock Comparison of skin stretch and vibrotactile stimulation for feedback of proprioceptive information.
\newblock In {\em 2008 Symposium on haptic interfaces for virtual environment and teleoperator systems}, pp. 71--78. IEEE, 2008.

\bibitem{blalock2010encoding}
L.~D. Blalock and B.~A. Clegg.
\newblock Encoding and representation of simultaneous and sequential arrays in visuospatial working memory.
\newblock {\em Quarterly Journal of Experimental Psychology}, 63(5):856--862, 2010.

\bibitem{blauert1997spatial}
J.~Blauert.
\newblock {\em Spatial hearing: the psychophysics of human sound localization}.
\newblock MIT press, 1997.

\bibitem{on_head_vib_display}
T.~Chen, Y.-S. Wu, and K.~Zhu.
\newblock Investigating different modalities of directional cues for multi-task visual-searching scenario in virtual reality.
\newblock In {\em Proceedings of the 24th ACM Symposium on Virtual Reality Software and Technology}, VRST '18. Association for Computing Machinery, New York, NY, USA, 2018. doi: {{%
10\hspace{.1pt}\discretionary{.}{%
}{.}\hspace{.4pt}1145\discretionary{/}{%
}{/}3281505\hspace{.1pt}\discretionary{.}{%
}{.}\hspace{.4pt}3281516}}


\bibitem{collignon2011functional}
O.~Collignon, G.~Vandewalle, P.~Voss, G.~Albouy, G.~Charbonneau, M.~Lassonde, and F.~Lepore.
\newblock Functional specialization for auditory--spatial processing in the occipital cortex of congenitally blind humans.
\newblock {\em Proceedings of the National Academy of Sciences}, 108(11):4435--4440, 2011.

\bibitem{collignon2009cross}
O.~Collignon, P.~Voss, M.~Lassonde, and F.~Lepore.
\newblock Cross-modal plasticity for the spatial processing of sounds in visually deprived subjects.
\newblock {\em Experimental brain research}, 192:343--358, 2009.

\bibitem{collins2005cutaneous}
D.~F. Collins, K.~M. Refshauge, G.~Todd, and S.~C. Gandevia.
\newblock Cutaneous receptors contribute to kinesthesia at the index finger, elbow, and knee.
\newblock {\em Journal of neurophysiology}, 94(3):1699--1706, 2005.

\bibitem{doherty2024hapt}
C.~Doherty, J.~L. Tennison, and J.~L. Gorlewicz.
\newblock Hapt-x-pand: The design and evaluation of a radially expanding and contracting skin drag haptic device.
\newblock In {\em 2024 IEEE Haptics Symposium (HAPTICS)}, pp. 264--270. IEEE, 2024.

\bibitem{edin2001cutaneous}
B.~B. Edin.
\newblock Cutaneous afferents provide information about knee joint movements in humans.
\newblock {\em The Journal of physiology}, 531(1):289--297, 2001.

\bibitem{drag_tap_vib}
L.~Fang, T.~Zhu, E.~Pescara, Y.~Huang, Y.~Zhou, and M.~Beigl.
\newblock Dragtapvib:\&nbsp;an on-skin electromagnetic drag, tap, and vibration actuator for wearable computing.
\newblock In {\em Proceedings of the Augmented Humans International Conference 2022}, AHs '22, p. 203–211. Association for Computing Machinery, New York, NY, USA, 2022. doi: {{%
10\hspace{.1pt}\discretionary{.}{%
}{.}\hspace{.4pt}1145\discretionary{/}{%
}{/}3519391\hspace{.1pt}\discretionary{.}{%
}{.}\hspace{.4pt}3519395}}


\bibitem{vibrotactile_survey}
P.~Galambos.
\newblock Vibrotactile feedback for haptics and telemanipulation: Survey, concept and experiment.
\newblock {\em Acta Polytechnica Hungarica}, 9(1):41--65, 2012.

\bibitem{gallace2009cognitive}
A.~Gallace and C.~Spence.
\newblock The cognitive and neural correlates of tactile memory.
\newblock {\em Psychological bulletin}, 135(3):380, 2009.

\bibitem{zelda}
D.~Gon\c{c}alves, M.~Pi\c{c}arra, P.~Pais, J.~a. Guerreiro, and A.~Rodrigues.
\newblock "my zelda cane": Strategies used by blind players to play visual-centric digital games.
\newblock In {\em Proceedings of the 2023 CHI Conference on Human Factors in Computing Systems}, CHI '23. Association for Computing Machinery, New York, NY, USA, 2023. doi: {{%
10\hspace{.1pt}\discretionary{.}{%
}{.}\hspace{.4pt}1145\discretionary{/}{%
}{/}3544548\hspace{.1pt}\discretionary{.}{%
}{.}\hspace{.4pt}3580702}}


\bibitem{mediapipe}
Google.
\newblock Mediapipe, 2023.
\newblock Accessed: Sept. 16, 2024.

\bibitem{gougoux2005functional}
F.~Gougoux, R.~J. Zatorre, M.~Lassonde, P.~Voss, and F.~Lepore.
\newblock A functional neuroimaging study of sound localization: visual cortex activity predicts performance in early-blind individuals.
\newblock {\em PLoS biology}, 3(2):e27, 2005.

\bibitem{TactileGlove}
S.~G\"{u}nther, F.~M\"{u}ller, M.~Funk, J.~Kirchner, N.~Dezfuli, and M.~M\"{u}hlh\"{a}user.
\newblock Tactileglove: Assistive spatial guidance in 3d space through vibrotactile navigation.
\newblock In {\em Proceedings of the 11th PErvasive Technologies Related to Assistive Environments Conference}, PETRA '18, p. 273–280. Association for Computing Machinery, New York, NY, USA, 2018. doi: {{%
10\hspace{.1pt}\discretionary{.}{%
}{.}\hspace{.4pt}1145\discretionary{/}{%
}{/}3197768\hspace{.1pt}\discretionary{.}{%
}{.}\hspace{.4pt}3197785}}


\bibitem{nasa_tlx}
S.~G. Hart and L.~E. Staveland.
\newblock Development of nasa-tlx (task load index): Results of empirical and theoretical research.
\newblock In {\em Advances in psychology}, vol.~52, pp. 139--183. Elsevier, 1988.

\bibitem{hayward2000tactile}
V.~Hayward and J.~M. Cruz-Hernandez.
\newblock Tactile display device using distributed lateral skin stretch.
\newblock In {\em ASME international mechanical engineering congress and exposition}, vol. 26652, pp. 1309--1314. American Society of Mechanical Engineers, 2000.

\bibitem{hong2017evaluating}
J.~Hong, A.~Pradhan, J.~E. Froehlich, and L.~Findlater.
\newblock Evaluating wrist-based haptic feedback for non-visual target finding and path tracing on a 2d surface.
\newblock In {\em Proceedings of the 19th International ACM SIGACCESS Conference on Computers and Accessibility}, pp. 210--219, 2017.

\bibitem{iachini2014does}
T.~Iachini, G.~Ruggiero, and F.~Ruotolo.
\newblock Does blindness affect egocentric and allocentric frames of reference in small and large scale spaces?
\newblock {\em Behavioural brain research}, 273:73--81, 2014.

\bibitem{ion}
A.~Ion, E.~J. Wang, and P.~Baudisch.
\newblock Skin drag displays: Dragging a physical tactor across the user's skin produces a stronger tactile stimulus than vibrotactile.
\newblock In {\em Proceedings of the 33rd Annual ACM Conference on Human Factors in Computing Systems}, CHI '15, p. 2501–2504. Association for Computing Machinery, New York, NY, USA, 2015. doi: {{%
10\hspace{.1pt}\discretionary{.}{%
}{.}\hspace{.4pt}1145\discretionary{/}{%
}{/}2702123\hspace{.1pt}\discretionary{.}{%
}{.}\hspace{.4pt}2702459}}


\bibitem{je_design}
S.~Je, O.~Choi, K.~Choi, M.~Lee, H.-J. Suk, L.~Chan, and A.~Bianchi.
\newblock Designing skin-dragging haptic motions for wearables.
\newblock In {\em Proceedings of the 2017 ACM International Symposium on Wearable Computers}, ISWC '17, p. 98–101. Association for Computing Machinery, New York, NY, USA, 2017. doi: {{%
10\hspace{.1pt}\discretionary{.}{%
}{.}\hspace{.4pt}1145\discretionary{/}{%
}{/}3123021\hspace{.1pt}\discretionary{.}{%
}{.}\hspace{.4pt}3123050}}


\bibitem{TactoRing}
S.~Je, B.~Rooney, L.~Chan, and A.~Bianchi.
\newblock Tactoring: A skin-drag discrete display.
\newblock In {\em Proceedings of the 2017 CHI Conference on Human Factors in Computing Systems}, CHI '17, p. 3106–3114. Association for Computing Machinery, New York, NY, USA, 2017. doi: {{%
10\hspace{.1pt}\discretionary{.}{%
}{.}\hspace{.4pt}1145\discretionary{/}{%
}{/}3025453\hspace{.1pt}\discretionary{.}{%
}{.}\hspace{.4pt}3025703}}


\bibitem{vrbubble}
T.~F. Ji, B.~Cochran, and Y.~Zhao.
\newblock Vrbubble: Enhancing peripheral awareness of avatars for people with visual impairments in social virtual reality.
\newblock In {\em Proceedings of the 24th International ACM SIGACCESS Conference on Computers and Accessibility}, ASSETS '22. Association for Computing Machinery, New York, NY, USA, 2022. doi: {{%
10\hspace{.1pt}\discretionary{.}{%
}{.}\hspace{.4pt}1145\discretionary{/}{%
}{/}3517428\hspace{.1pt}\discretionary{.}{%
}{.}\hspace{.4pt}3544821}}


\bibitem{lips_non-contact_tactile}
A.~Jingu, T.~Kamigaki, M.~Fujiwara, Y.~Makino, and H.~Shinoda.
\newblock Lipnotif: Use of lips as a non-contact tactile notification interface based on ultrasonic tactile presentation.
\newblock In {\em The 34th Annual ACM Symposium on User Interface Software and Technology}, UIST '21, p. 13–23. Association for Computing Machinery, New York, NY, USA, 2021. doi: {{%
10\hspace{.1pt}\discretionary{.}{%
}{.}\hspace{.4pt}1145\discretionary{/}{%
}{/}3472749\hspace{.1pt}\discretionary{.}{%
}{.}\hspace{.4pt}3474732}}


\bibitem{kent2023biomechanically}
J.~A. Kent.
\newblock Biomechanically-consistent skin stretch as an intuitive mechanism for sensory feedback: A preliminary investigation in the lower limb.
\newblock {\em IEEE Transactions on Haptics}, 16(1):106--111, 2023.

\bibitem{phantom_senstion_wearables}
J.~Kim, S.~Oh, C.~Park, and S.~Choi.
\newblock Body-penetrating tactile phantom sensations.
\newblock In {\em Proceedings of the 2020 CHI Conference on Human Factors in Computing Systems}, CHI '20, p. 1–13. Association for Computing Machinery, New York, NY, USA, 2020. doi: {{%
10\hspace{.1pt}\discretionary{.}{%
}{.}\hspace{.4pt}1145\discretionary{/}{%
}{/}3313831\hspace{.1pt}\discretionary{.}{%
}{.}\hspace{.4pt}3376619}}


\bibitem{SpinOcchio}
M.~J. Kim, N.~Ryu, W.~Chang, M.~Pahud, M.~Sinclair, and A.~Bianchi.
\newblock Spinocchio: Understanding haptic-visual congruency of skin-slip in vr with a dynamic grip controller.
\newblock In {\em Proceedings of the 2022 CHI Conference on Human Factors in Computing Systems}, CHI '22. Association for Computing Machinery, New York, NY, USA, 2022. doi: {{%
10\hspace{.1pt}\discretionary{.}{%
}{.}\hspace{.4pt}1145\discretionary{/}{%
}{/}3491102\hspace{.1pt}\discretionary{.}{%
}{.}\hspace{.4pt}3517724}}


\bibitem{low-cost_disability_device}
J.~L. K\"{o}nig, J.~Penaredondo, E.~McCullagh, J.~Bowen, and A.~Hinze.
\newblock Let's make it accessible: The challenges of working with low-cost commercially available wearable devices.
\newblock In {\em Proceedings of the 35th Australian Computer-Human Interaction Conference}, OzCHI '23, p. 493–503. Association for Computing Machinery, New York, NY, USA, 2024. doi: {{%
10\hspace{.1pt}\discretionary{.}{%
}{.}\hspace{.4pt}1145\discretionary{/}{%
}{/}3638380\hspace{.1pt}\discretionary{.}{%
}{.}\hspace{.4pt}3638415}}


\bibitem{lederman2011tactile}
S.~J. Lederman and L.~A. Jones.
\newblock Tactile and haptic illusions.
\newblock {\em IEEE Transactions on Haptics}, 4(4):273--294, 2011.

\bibitem{dynamic_tactile}
V.~Lehtinen, A.~Oulasvirta, A.~Salovaara, and P.~Nurmi.
\newblock Dynamic tactile guidance for visual search tasks.
\newblock In {\em Proceedings of the 25th Annual ACM Symposium on User Interface Software and Technology}, UIST '12, p. 445–452. Association for Computing Machinery, New York, NY, USA, 2012. doi: {{%
10\hspace{.1pt}\discretionary{.}{%
}{.}\hspace{.4pt}1145\discretionary{/}{%
}{/}2380116\hspace{.1pt}\discretionary{.}{%
}{.}\hspace{.4pt}2380173}}


\bibitem{leslie_2022_disabilitydongle}
Z.~Leslie.
\newblock The disability dongle: The limits of enabling technology, 2022.
\newblock Accessed: 2024-09-18.

\bibitem{miyakami2022head}
M.~Miyakami, A.~Takahashi, and H.~Kajimoto.
\newblock Head rotation and illusory force sensation by lateral skin stretch on the face.
\newblock {\em Frontiers in Virtual Reality}, 3:930848, 2022.

\bibitem{accessible_design}
M.~Mott, E.~Cutrell, M.~Gonzalez~Franco, C.~Holz, E.~Ofek, R.~Stoakley, and M.~Ringel~Morris.
\newblock Accessible by design: An opportunity for virtual reality.
\newblock In {\em 2019 IEEE International Symposium on Mixed and Augmented Reality Adjunct (ISMAR-Adjunct)}, pp. 451--454, 2019. doi: {{%
10\hspace{.1pt}\discretionary{.}{%
}{.}\hspace{.4pt}1109\discretionary{/}{%
}{/}ISMAR\discretionary{%
}{-}{-}Adjunct\hspace{.1pt}\discretionary{.}{%
}{.}\hspace{.4pt}2019\hspace{.1pt}\discretionary{.}{%
}{.}\hspace{.4pt}00122}}


\bibitem{haptiX}
M.~Pascher, T.~Franzen, K.~Kronhardt, U.~Gruenefeld, S.~Schneegass, and J.~Gerken.
\newblock Haptix: Vibrotactile haptic feedback for communication of 3d directional cues.
\newblock In {\em Extended Abstracts of the 2023 CHI Conference on Human Factors in Computing Systems}, CHI EA '23. Association for Computing Machinery, New York, NY, USA, 2023. doi: {{%
10\hspace{.1pt}\discretionary{.}{%
}{.}\hspace{.4pt}1145\discretionary{/}{%
}{/}3544549\hspace{.1pt}\discretionary{.}{%
}{.}\hspace{.4pt}3585601}}


\bibitem{renier2010preserved}
L.~A. Renier, I.~Anurova, A.~G. De~Volder, S.~Carlson, J.~VanMeter, and J.~P. Rauschecker.
\newblock Preserved functional specialization for spatial processing in the middle occipital gyrus of the early blind.
\newblock {\em Neuron}, 68(1):138--148, 2010.

\bibitem{SCHMIDT201343}
S.~Schmidt, C.~Tinti, M.~Fantino, I.~C. Mammarella, and C.~Cornoldi.
\newblock Spatial representations in blind people: The role of strategies and mobility skills.
\newblock {\em Acta Psychologica}, 142(1):43--50, 2013. doi: {{%
10\hspace{.1pt}\discretionary{.}{%
}{.}\hspace{.4pt}1016\discretionary{/}{%
}{/}j\hspace{.1pt}\discretionary{.}{%
}{.}\hspace{.4pt}actpsy\hspace{.1pt}\discretionary{.}{%
}{.}\hspace{.4pt}2012\hspace{.1pt}\discretionary{.}{%
}{.}\hspace{.4pt}11\hspace{.1pt}\discretionary{.}{%
}{.}\hspace{.4pt}010}}


\bibitem{schorr2013sensory}
S.~B. Schorr, Z.~F. Quek, R.~Y. Romano, I.~Nisky, W.~R. Provancher, and A.~M. Okamura.
\newblock Sensory substitution via cutaneous skin stretch feedback.
\newblock In {\em 2013 IEEE International Conference on Robotics and Automation}, pp. 2341--2346. IEEE, 2013.

\bibitem{shim}
Y.~Shim, T.~Kim, S.~Lee, S.~Kim, and G.~Lee.
\newblock Quadstretch: A forearm-wearable skin stretch display for immersive vr experience.
\newblock In {\em SIGGRAPH Asia 2022 Emerging Technologies}, SA '22. Association for Computing Machinery, New York, NY, USA, 2022. doi: {{%
10\hspace{.1pt}\discretionary{.}{%
}{.}\hspace{.4pt}1145\discretionary{/}{%
}{/}3550471\hspace{.1pt}\discretionary{.}{%
}{.}\hspace{.4pt}3564761}}


\bibitem{squeri2012two}
V.~Squeri, A.~Sciutti, M.~Gori, L.~Masia, G.~Sandini, and J.~Konczak.
\newblock Two hands, one perception: how bimanual haptic information is combined by the brain.
\newblock {\em Journal of Neurophysiology}, 107(2):544--550, 2012.

\bibitem{tan2003haptic}
H.~Tan, R.~Gray, J.~J. Young, and R.~Taylor.
\newblock A haptic back display for attentional and directional cueing.
\newblock 2003.

\bibitem{tsukada2004activebelt}
K.~Tsukada and M.~Yasumura.
\newblock Activebelt: Belt-type wearable tactile display for directional navigation.
\newblock In {\em international conference on ubiquitous computing}, pp. 384--399. Springer, 2004.

\bibitem{skin_stretch_on_face}
C.~Wang, D.-Y. Huang, S.-w. Hsu, C.-E. Hou, Y.-L. Chiu, R.-C. Chang, J.-Y. Lo, and B.-Y. Chen.
\newblock Masque: Exploring lateral skin stretch feedback on the face with head-mounted displays.
\newblock In {\em Proceedings of the 32nd Annual ACM Symposium on User Interface Software and Technology}, UIST '19, p. 439–451. Association for Computing Machinery, New York, NY, USA, 2019. doi: {{%
10\hspace{.1pt}\discretionary{.}{%
}{.}\hspace{.4pt}1145\discretionary{/}{%
}{/}3332165\hspace{.1pt}\discretionary{.}{%
}{.}\hspace{.4pt}3347898}}


\bibitem{skin_stretch_on_leg}
C.~Wang, D.-Y. Huang, S.-W. Hsu, C.-L. Lin, Y.-L. Chiu, C.-E. Hou, and B.-Y. Chen.
\newblock Gaiters: Exploring skin stretch feedback on legs for enhancing virtual reality experiences.
\newblock In {\em Proceedings of the 2020 CHI Conference on Human Factors in Computing Systems}, CHI '20, p. 1–14. Association for Computing Machinery, New York, NY, USA, 2020. doi: {{%
10\hspace{.1pt}\discretionary{.}{%
}{.}\hspace{.4pt}1145\discretionary{/}{%
}{/}3313831\hspace{.1pt}\discretionary{.}{%
}{.}\hspace{.4pt}3376865}}


\bibitem{weber2011evaluation}
B.~Weber, S.~Sch{\"a}tzle, T.~Hulin, C.~Preusche, and B.~Deml.
\newblock Evaluation of a vibrotactile feedback device for spatial guidance.
\newblock In {\em 2011 IEEE World Haptics Conference}, pp. 349--354. IEEE, 2011.

\bibitem{spaceSense}
K.~Yatani, N.~Banovic, and K.~Truong.
\newblock Spacesense: Representing geographical information to visually impaired people using spatial tactile feedback.
\newblock In {\em Proceedings of the SIGCHI Conference on Human Factors in Computing Systems}, CHI '12, p. 415–424. Association for Computing Machinery, New York, NY, USA, 2012. doi: {{%
10\hspace{.1pt}\discretionary{.}{%
}{.}\hspace{.4pt}1145\discretionary{/}{%
}{/}2207676\hspace{.1pt}\discretionary{.}{%
}{.}\hspace{.4pt}2207734}}


\bibitem{vibhand}
K.~Zhao, M.~Serrano, B.~Oriola, and C.~Jouffrais.
\newblock Vibhand: On-hand vibrotactile interface enhancing non-visual exploration of digital graphics.
\newblock {\em Proceedings of the ACM on Human-Computer Interaction}, 4(ISS):1--19, 2020.

\end{thebibliography}
\end{document}